\documentclass[12pt,letterpaper,hyper]{JHEP3}
\usepackage{amsmath,epsfig}
\usepackage{amssymb,amsfonts}
\usepackage{multirow,cite}
\usepackage{subfigure}
\usepackage{enumerate}

\newcommand{\bi}{\begin{itemize}}
\newcommand{\ei}{\end{itemize}}
\newcommand{\non}{\nonumber}
\def\p{\partial}
\def\a{\alpha}
\def\b{\beta}
\def\d{\delta}
\def\g{\gamma}
\def\l{\lambda}
\def\e{\epsilon}
\def\k{\kappa}

\def\om{\omega}

\def\O{\mathcal{O}}

\def\L{\Lambda}

\def\r{\rightarrow}
\def\half{{\frac12}}

\newcommand{\bea}{\begin{eqnarray}}
\newcommand{\eea}{\end{eqnarray}}
\newcommand{\be}{\begin{equation}}
\newcommand{\ee}{\end{equation}}
\newcommand{\re}[1]{(\ref{#1})}

\def\sst#1{{\scriptscriptstyle #1}}

\title{The Curious Case of Null Warped Space}
\author{Dionysios Anninos$^{\heartsuit }$, Geoffrey Comp\`ere$^{\flat}$, Sophie de Buyl$^{\flat}$,
 St\'ephane Detournay$^{\natural}$ and Monica Guica$^{\dagger}$\vspace{0.25 cm}\\
$^{\heartsuit}$ Center for the Fundamental Laws of Nature, Harvard University,\\
\hspace*{0.15cm}Cambridge, MA 02138, USA \vspace{0.25 cm}
 \\
$^{\flat}$ Department of Physics, University of California, \\
\hspace*{0.15cm} Santa Barbara, CA 93106, USA \vspace{0.25 cm}\\
 $^{\natural}$ Kavli Institute for Theoretical Physics, University of California,\\
\hspace*{0.15cm} Santa Barbara, CA 93106, USA \vspace{0.25 cm} \\
$^{\dagger}$
Laboratoire de Physique Th\'eorique et Hautes Energies(LPTHE) \\
\hspace*{0.15cm} Tour24-25, 5 \`eme \'etage, Boite 126 , 4 Place Jussieu  \\
\hspace*{0.15cm} 75252 Paris Cedex 05, France
}

\preprint{ NSF-KITP-10-020}

\abstract{We initiate a comprehensive study of a set of solutions of topologically massive gravity known as null warped anti-de Sitter spacetimes. These are pp-wave extensions of three-dimensional anti-de Sitter space. We first perform a careful analysis
of the linearized stability of black holes in these spacetimes. We find two qualitatively different types of solutions to the linearized equations of motion: the first set has an exponential time dependence, the second - a polynomial
time dependence. The solutions polynomial in time induce severe pathologies and moreover  survive at the non-linear level. In order to make sense of these geometries, it is thus
crucial to impose appropriate boundary conditions. We argue that there
exists a consistent set of boundary conditions that allows us to reject the above pathological modes
from the physical spectrum. The asymptotic symmetry group associated to these boundary conditions consists of a centrally-extended Virasoro algebra.
Using this central charge we can account for the entropy of the black holes via Cardy's formula. Finally, we note that the black hole spectrum is chiral and prove a Birkoff
theorem showing that there are no other stationary axisymmetric black holes with the specified
asymptotics. We extend most of the analysis to a larger family of pp-wave black holes which are related to Schr\"odinger spacetimes with critical exponent $z$.}

\begin{document}

\section{Introduction and summary}

Gaining control over questions in quantum gravity remains a deep challenge in theoretical physics. String theory has provided important insights, however it contains an intricate and at times overwhelming structure. A complementary avenue is given by studying theories of pure gravity in three-dimensional spacetime. Their advantage comes in the form of their simplicity: they have fewer degrees of freedom and  much simpler dynamics than their higher-dimensional cousins, while still allowing for interesting black hole solutions.

Of particular interest has been the proposal of \cite{Witten:2007kt} - that theories of pure gravity in AdS$_3$ should be holographically dual to two-dimensional extremal conformal field theories. Realizing this proposal for Einstein gravity in three-dimensions has encountered various conceptual difficulties \cite{Maloney:2007ud}. These difficulties were nevertheless surmounted for the case of a different theory of pure gravity in three dimensions known as chiral gravity \cite{Li:2008dq}, whose partition function in AdS$_3$ was shown to precisely take the form of the partition function of an extremal conformal field theory \cite{Maloney:2009ck}. If such conformal field theories are proven to exist, this will be the first example of a well defined theory of quantum gravity with only metric degrees of freedom.

Chiral gravity is a particular case of the theory of topologically massive gravity (TMG) \cite{Deser:1981wh}, which contains a gravitational Chern-Simons term in addition to the usual Einstein-Hilbert action. This theory has a stable anti-de Sitter vacuum at a special point in the parameter space known as the chiral point. The chiral point is given by $\mu \ell = 1$ where $\mu$ is the Chern-Simons coefficient and $-1/\ell^2$ is the cosmological constant.

If it is true that TMG makes sense at a single point in parameter space, it is natural to ask whether the theory contains other consistent vacua for different values of the coupling, that is for $\mu \ell \neq 1$. It turns out that for $\mu \ell > 3$ TMG has a set of stable vacua known as spacelike warped AdS$_3$ - spacetimes with isometry group $SL(2,\mathbb{R})_R \times U(1)_L$ and intriguing asymptotic structure different from that of AdS$_3$. These spacetimes contain a rich black hole spectrum, whose entropy  can be reproduced by a Cardy-like formula involving both left and right-movers. This match is quite surprising, given the fact that the asymptotic symmetry group of these spacetimes consist of only one centrally extended Virasoro algebra and a current algebra \cite{Compere:2008cv,Compere:2009zj,Blagojevic:2009ek}.

At $\mu \ell=3$, TMG has another interesting vacuum solution which is a pp-wave extension of three-dimensional anti-de Sitter space and is known as null warped AdS$_3$. This spacetime has a null Killing vector and is the three-dimensional analogue (up to identifications) of the Schr\"odinger spacetimes that have been proposed as holographic duals to non-relativistic CFTs \cite{Son:2008ye}. This spacetime also contains a rich chiral black hole spectrum, known as null warped black holes. Our main focus in this article is to give a detailed account of the physics of these objects.

The first issue we address is that of stability. For this, we perform an extensive study of the linearized equations of motion about an arbitrary null warped black hole at $\mu\ell = 3$. We find two types of solutions. The first set of solutions are oscillatory in time and have the general feature that their asymptotic falloff depends on their (possibly complex) frequency $\omega$. This is reminiscent of the behavior of modes in spacelike warped anti-de Sitter space where the falloff is momentum dependent. In particular, for frequencies with a large enough real part the modes acquire a non-trivial flux across the boundary at large radius and are interpreted as travelling waves. As in the case of spacelike warped AdS$_3$ and chiral gravity, the spacetime can be stable only if appropriate boundary conditions are imposed, which exclude the oscillating modes.

The second set of solutions we uncover have polynomial time dependence and are severely pathological. In particular, for large times they may lead to the development of closed timelike curves in the geometry, having started from perfectly regular initial data. Fortunately, we show that these modes can be decoupled at the nonlinear level by an appropriate choice of initial data and a (more restrictive) set of boundary conditions at infinity.

The two options for boundary conditions that we propose pass all the non-trivial checks of consistency - they give rise to finite, integrable and conserved charges. Only one of the two proposals leads us to discard the pathological polynomial modes. Due to our incomplete understanding of the organization of the non-linear equations of motion in asymptotically null warped spacetimes, we are unable to give a sharp physical interpretation of the proposed boundary conditions.

Furthermore, both boundary conditions give rise to a rich asymptotic symmetry algebra: a single copy of a centrally extended Virasoro algebra times (for the less restrictive boundary conditions only) a $\widehat{u(1)}$ current algebra with vanishing level. This finding suggests that TMG at $\mu \ell=3$ with the given boundary conditions may be dual to a (chiral half of a) conformal field theory with the respective central charge. As a consistency check, we successfully retrieve the entropy of the asymptotically null warped black holes using the (chiral) Cardy formula.

It is interesting to note that both the black hole spectrum and the spectrum of perturbations (with appropriate boundary conditions) is chiral. Given the success of chiral gravity, we further investigate the question of chirality of TMG in null warped AdS$_3$. To this end we prove a Birkoff-type theorem stating that the null warped black holes are the most general analytic time-independent axisymmetric solutions and that are asymptotically null warped. Despite a large amount of evidence in favor of chirality at the classical level, it is not clear whether the truncated theory is by itself consistent, so we leave this issue open.

We extend part of this analysis to a more general set of black holes of TMG, which are natural generalizations to $\mu \ell > 3$ of the null warped black holes. These spacetimes also possess a null Killing vector and exhibit anisotropic scaling at large radius, with a (spatial) dynamical exponent $z$. We therefore playfully refer to these solutions as null z-warped black holes. They belong to the family of Schr\"{o}dinger
geometries that has been extensively studied in the recent literature \cite{Balasubramanian:2008dm}.
Unlike anti-de Sitter space, these spacetimes do not have a well understood conformal boundary and there is no known Fefferman-Graham type expansion of the asympotically null z-warped solution space.

We find the equation for linearized gravitational perturbations around the z-warped black hole backgrounds, which again has solutions exponential and polynomial in time. The asymptotic behavior of the exponential modes is extremely different from the behavior of their counterparts in null warped or regular AdS$_3$ - in particular, they are all non-normalizable. We also propose  boundary conditions for null z-warped spacetimes, but for technical reasons leave the proof of their full consistency to future work. These boundary conditions resemble those used in Kerr/CFT \cite{Guica:2008mu}, and they yield a Virasoro asymptotic symmetry group with a central extension which precisely matches the z-warped black hole entropy via Cardy's formula.

Along the way, we uncover many  peculiarities of the null warped and z-warped spacetimes: solutions with hair, three-dimensional black holes that cannot be obtained via identifications of a `vacuum' geometry, smooth solitons and polynomial modes. In addition, we study the behavior of geodesics in null warped AdS$_3$ and develop techniques for proving Birkhoff-type theorems in TMG using the formalism of \cite{Clement:1994sb,Bouchareb:2007yx} and method of \cite{Maloney:2009ck}. All these potentially useful facts can be found in the multiple appendices.

The nature of this work has been explorative. There is relatively little known about the spacetimes we have studied, but they already seem to contain a rich amount of physics and they are beginning to play a role in various areas of research. We believe we have made a small step forward in understanding them. However, much is left to be done. For instance, it is crucial to obtain a better understanding of the asymptotic structure of these spacetimes, and how to regularize the on-shell action and compute the boundary stress tensor. Also, more robust arguments are required to show that the spectrum of propagating modes can be consistently truncated if we are to propose that TMG has consistent chiral null warped vacua. A better understanding of the boundary conditions in these spacetimes is essential.

The organization of this paper is as follows: in section \ref{frame} we describe the black hole solutions that we will be concerned with. In section \ref{perts} we solve the linearized equations of motion for both warped and z-warped backgrounds and study normalizability of the solutions. In section \ref{ASG} we discuss the various choices of boundary conditions for null warped spacetimes and prove their consistency. In section \ref{CardySec} we show that the central charge of the Virasoro part of the asymptotic symmetries along with the energy of the black hole reproduce the black hole entropy via Cardy's formula. We also comment on the chirality of the spectrum and the relation between null warped and spacelike warped black holes. In section \ref{ASGz} we propose a (tentative) set of boundary conditions for null z-warped spacetimes and show that Cardy's formula again works. %Section 7 is reserved for conclusions.

\section{Null z-warped black holes}\label{frame}

In this section we discuss the framework  we will be working with, namely topologically massive gravity with a negative cosmological constant, and the geometry of the solutions relevant to our discussion.

\subsection{Framework and solutions}

The theory we will be considering throughout the work is topologically massive gravity (TMG) \cite{Deser:1981wh,Deser:1982vy} whose action is given by
\be
I_{TMG} = \frac{1}{16\pi G}\int d^3x \sqrt{-g}\left[  R - 2\Lambda + \frac{1}{2\mu}\varepsilon^{\lambda \mu \nu} \Gamma^\rho_{\lambda \sigma} ( \partial_\mu \Gamma^\sigma_{\rho\nu} + \frac{2}{3} \Gamma^\sigma_{\mu\tau} \Gamma^{\tau}_{\nu \rho} ) \right].
\ee
Here $\Lambda = -1/\ell^2$ is the cosmological constant and $\mu$ is the Chern-Simons coefficient. The three-dimensional Newton constant $G$ is taken to be positive. The equations of motion for TMG with a cosmological constant are given by
\be
E_{\mu \nu} \equiv R_{\mu\nu} - \frac{1}{2}R \, g_{\mu\nu} - \frac{1}{\ell^2} g_{\mu\nu} + \frac{1}{\mu}C_{\mu\nu} = 0 \label{eomTMG}
\ee
where
\be
C_{\mu \nu} = \half \,{{\varepsilon_{\mu}}^{\alpha \beta}}\nabla_{\alpha}(R_{\beta\nu} - \frac{1}{4}g_{\beta\nu}R)
\ee
is the Cotton tensor and $\varepsilon^{\mu\nu\rho}$ is the Levi-Civita tensor. The above theory is known to have a rich set of vacua \cite{Nutku:1993eb,Gurses:2008wu,Anninos:2008fx,Chow:2009km,Anninos:2009jt} which have raised recent interest. Of these solutions, the ones we will be focusing on are known as the null warped AdS$_3$ black holes solutions, which we now proceed to describe.

These spacetimes generically possess two isometries, one generated by a null Killing vector $\p_t$ and the other by a spacelike compact\footnote{If the $U(1)$ is not compact then $\beta$ is pure gauge and can be set to zero. The case $\a=0$ corresponds to null warped AdS$_3$ in Poincar\'e coordinates, while $\a^2=-1$ corresponds to null warped AdS$_3$ in global coordinates, having an $SL(2,R) \times U(1)_{null}$ isometry group. %Null warped AdS$_3$ is actually part of an exact string background, through a parabolic asymmetric marginal deformation of the $SL(2,R)$ WZW model \cite{Israel:2004vv}.
} $U(1)$ Killing vector $\p_{\phi}$. The general metric of the solutions we will be studying is given by \cite{Clement:1994sb,Deser:2004wd,Gibbons:2008vi}

\be\label{metric}
\frac{ds^2}{\ell^2} = {2 r dt d\phi} + f(r) \, d\phi^2  + \frac{dr^2}{4r^2}, \hspace{1cm} \varepsilon_{r\phi t}=+\sqrt{-g},\ee
where\footnote{This spacetime can be thought of as the truncation to three dimensions of the double analytical continuation of the Schr\"odinger geometry \cite{Balasubramanian:2008dm} which interchanges time with a space direction. The number $z$ should therefore not be interpreted as a dynamical critical exponent. Indeed, in the absence of $\beta$, $\alpha$ and when $\phi$ is non-compact, the solution is invariant under the dilatation
\be
D \equiv -\frac{2}{z} r \frac{\p}{\p r} + \phi \frac{\p}{\p \phi} + \frac{2-z}{z} t \frac{\p}{\p t}.
\ee
The scaling of (null) time with respect to the spatial $\phi$ coordinate is $Z = \frac{2-z}{z}$, which is better interpreted as a critical exponent.}
\be
f(r)= r^{z} + \beta \, r + \alpha^2 \;, \;\;\;\;\;z \equiv \frac{\mu \ell +1}{2} \equiv \frac{3 \nu+1}{2}.\label{NullBH}
\ee
Here $\phi$ is a periodic coordinate, $\phi \sim \phi + 2 \pi$. The function $f(r)$ is required to not have any positive roots, which in turn implies that $\beta \geq 0$. The ranges of the coordinates are  $t \in (-\infty,\infty)$ and $r \in (r_s, \infty )$, where $r_s$ is the largest (negative) real root of $f(r)$. In appendix \ref{nullgeom} we discuss the geometry of these spacetimes, including the case when $\phi$ is non-compact.

All the above spacetimes have an asymptotic boundary at $r = \infty$. Depending on the value of $z$, we can distinguish several types of asymptotic behavior
\bi
\item for  $z = 2$, the solution will be referred to as being asymptotically \emph{null warped} AdS$_3$ \item for  $z > 2$, the solution will be referred to as being asymptotically \emph{null z-warped} AdS$_3$
\item if $z \leq 1$ then the metric \eqref{NullBH} becomes asymptotically AdS$_3$ under the standard Brown-Henneaux boundary conditions \cite{Brown:1986nw} or anti-de Sitter boundary conditions in TMG \cite{Hotta:2008yq,Henneaux:2009pw}.
\ei
In the range $1< z <2$, the above spacetimes exhibit infinite tidal forces \cite{Blau:2009gd} near $r=0$; therefore we will focus on the case $z \geq 2$. All the above spacetimes have a horizon at $r=0$, which is of finite size due to the identification of $\phi$. When the function $f(r)$ becomes negative the solution has closed timelike curves, which are necessarily inside the horizon $(r<0)$. Due to such causal singularities, these solutions are termed black holes.%\footnote{The solution with $\alpha=0$ contains a closed null curve at $r=0$ in analogy with the zero mass BTZ black hole.}

%Notice that in the $(1,0)$ supersymmetric extension of TMG, all solutions \eqref{NullBH} have one Killing spinor \cite{Chow:2009km}.

\subsection{Null warped  black holes and smooth solitons}

When $z=2$ and $\beta \geq 2 \a$ the solution \eqref{NullBH} describes the so-called null warped black holes. They have a causal singularity hidden by the horizon and they can be obtained as discrete global identifications of the `vacuum' null warped geometry with $\alpha = \beta = 0$ and $\phi$ unidentified.

These black holes have two Killing vectors given by $\partial_t$ and $\partial_\phi$ from which we can obtain the conserved ADT charges
\be\label{adtcharge}
Q_{\partial_t} = 0,  \quad Q_{\partial_\phi} =  \frac{{\alpha}^2 \ell}{3G}.
\ee
The Chern-Simons corrected entropy can be computed and is found to be
\be
S_{BH} = \frac{2 \pi {\alpha} \ell }{3 G}.\label{nullent}
\ee
In section \ref{ASG} we will obtain the above entropy \eqref{nullent} by considering the central extension of the Virasoro algebra and applying  Cardy's formula.

The surface gravity of these black holes vanishes and consequently so does their Hawking temperature. However, a ``right-moving'' temperature can be defined starting from the first law
\be\label{first}
\d S_{BH} = \frac{1}{T_R} \d Q_{\partial_\phi}.
\ee
The existence of causal singularities behind the horizon places an upper bound on the right-moving temperature
\be
T_{R} \leq T_{R}^{Max},\qquad  T_R = \frac{\alpha}{\pi l},\qquad T_{R}^{Max} \equiv \frac{\beta}{2 \pi l}.\label{temperatmax}
\ee
The bound \eqref{temperatmax} may look strange. As we already mentioned, it originates from the requirement that $f(r)$ have a zero somewhere for negative $r$. If we relax this requirement, the corresponding solutions still have a horizon at $r=0$, but which no longer hides a causal singularity. We call these solutions ``smooth black solitons''. All the formulae above in this section, including the first law of thermodynamics, the expression for the entropy and right-moving temperature, still hold in this range of parameter space.%\emph{Can the smooth solitons be obtained by identifications of null warped?}

The near-horizon geometry of the black hole spacetimes \re{NullBH} (which necessarily have $\alpha \neq 0$) is given by
\begin{equation}\label{NHG}
\frac{ds^2}{\ell^2} = \frac{1}{4}\left(-\frac{4r^2}{\alpha^2} dt^2 + \frac{dr^2}{r^2}\right) + \left( \alpha d\phi - \frac{r}{\alpha} dt\right)^2.
\end{equation}
This geometry is the well-known self-dual orbifold of AdS$_3$ \cite{Coussaert:1994tu}. The identification of this orbifold corresponds precisely to
\be
T_R = \frac{\a}{\pi \ell} \label{tright}.
\ee

\subsection{Null z-warped black holes $(z>2)$}

The null z-warped black holes, i.e. \eqref{NullBH} with $z>2$,  \emph{cannot} be obtained from an identification
of the `vacuum' null z-warped geometry $\alpha=\beta=0$. A proof of this is presented in appendix \ref{proof}.
Nevertheless, all local curvature invariants of the black hole spacetimes are identical to those of the ``vacuum''.
Indeed, the tensor $R^\mu_{\;\;\nu}$ is independent of $\alpha$ and $\beta$ and so are all
contractions of the Ricci tensor. %We checked that all curvature invariants that include the first derivative
%the Ricci are identical for the black holes and the background as well.
The property that physical parameters are not captured by curvature invariants is not unusual for pp-wave spacetimes, see e.g. \cite{Pravda:2002us} and is related to the fact that these metrics belong to the class of Kundt metrics \cite{Coley:2009eb, Coley:2007ib, Coley:2010ze, Coley:2009ut}.
Once again, the axisymmetric null z-warped black holes have two isometries given by $\partial_t$ and $\partial_\phi$ with corresponding conserved ADT charges
\be\label{Qphiz}
Q_{\partial_t} = 0, \quad Q_{\partial_\phi} = \frac{ \alpha^2 (\mu\ell + 1)}{4\mu G}.
\ee
We can further compute the Chern-Simons corrected entropy and the right-moving temperature of the axisymmetric z-warped black holes to find
\be
S_{BH} = \frac{\pi \alpha (\mu\ell + 1)}{ 2 G \mu\ell}, \qquad  T_R = \frac{\alpha}{\pi l}.\label{zBHentropy}
\ee
The conserved charges and entropy obey the first law of thermodynamics \eqref{first}. In section \ref{ASGz} we will discuss an Ansatz for the asymptotic symmetries of the z-warped solutions which consists of a Virasoro with a central extension that reproduces the black hole entropy.

%As discussed in \cite{Maldacena:1998bw, Anninos:2008fx,Balasubramanian:2009bg}, the self-dual $AdS_3$ orbifold can be viewed as a thermal state in the putative dual CFT, with vanishing right moving temperature.

The near horizon geometry of the null z-warped black holes is again \eqref{NHG}. The self-dual orbifold of AdS$_3$ with \eqref{tright} is therefore the universal near-horizon limit for all null warped and z-warped black holes.\footnote{It is amusing to note that the asymptotic symmetries of the near-horizon self-dual AdS$_3$ consist of a single copy of the Virasoro algebra whose central extension is $c_R = \frac{3l}{2G}(1+\frac{1}{\mu l})$. This central charge and the temperature \eqref{tright} allow one to reproduce the black hole entropy via the (chiral) Cardy formula. We note, however, that the stability of the self-dual quotient of AdS$_3$ in TMG has yet to be analyzed. }

\subsection{Hairy null z-warped black holes}

We would like to parenthetically mention that TMG also possesses a set of non-axisymmetric null z-warped black hole solutions, with metric of the form \eqref{NullBH}, but with
\be
f(r) = r^z + \b \, r + \a( \phi)^2 \;, \;\;\;\;\; z>2
\ee
where $\a(\phi)$ is an arbitrary periodic function. The conserved charge associated to the (asymptotic) symmetry $\p_{\phi}$ is
\be
Q_{\p_{\phi}} =  {\mu \ell + 1\over 8 \pi \mu \, G} \int^{2\pi}_0 \alpha(\phi)^2\, d \phi.
\ee
We call these solutions ``hairy black holes'' because we have an infinite set of solutions for fixed asymptotics and conserved charge. When $z=2$ the hairs can be removed by globally-defined diffeomorphisms. For $z>2$  this is no longer the case, the hairs are physical, and the solution space is consequently much richer.

\section{Perturbations of null z-warped AdS$_3$}\label{perts}

The null z-warped geometries have some unusual properties which render the dynamical analysis of linear fields intricate. First, these spacetimes do not have a timelike Killing vector, even locally. They rather have a global null Killing vector $\partial_t$ which we will refer to as time, since hypersurfaces of constant $t$ are spacelike. Second, these spacetimes do not have a Cauchy surface, which implies that the initial value problem is ill-posed without specifying the boundary conditions of the linear fields (see \cite{Ishibashi:2004wx} for an analysis of the dynamics of linear fields in anti-de Sitter space).

In this section we will study linear perturbations around the asymptotically warped and z-warped black holes. This will allow us to probe the behavior of the propagating modes of TMG and study the stability of these black holes. The asymptotic behavior of the solutions will provide insight towards defining boundary conditions for asymptotically null z-warped spacetimes.

This section is organized in three parts. In the first part we find all the solutions of the linear differential equation that the gauge-fixed perturbations satisfy. In the second part, we analyze the physical properties of the solutions about the $z=2$ background, most notably normalizability. Finally, we analyze the linearized solutions about the $z>2$ backgrounds in the third part.

\section*{{Part I: General linear analysis}}

We begin with the metric
\be
\frac{ds^2}{\ell^2} = \bar g_{\mu\nu} dx^\mu dx^\nu=  2 r d \phi dt + \left(  r^{z} + \beta  r + \alpha^2  \right) d \phi ^2  + \frac{dr^2}{4r^2}\label{eq:98}
\ee
with $\alpha >0$ and solve the linearized TMG equations for a perturbation $h_{\mu\nu}$ in synchronous gauge
\be
h_{t\mu} =0. \label{syncg}
\ee
which can always be reached. Since the $\phi$ direction is periodic, we can express the perturbations in a Fourier series:
\be
h_{\mu\nu} (r,t,\phi) = \sum_{m} h_{\mu\nu}^m (r,t) \, e^{- i m \phi}
\ee
with
\be
h_{\mu\nu}^m (r,t) = \left(\begin{array}{ccc} g_1^m(r,t) &  g_2^m(r,t) &  0  \\ g_2^m(r,t) & g_3^m(r,t) & 0 \\ 0 & 0 & 0\end{array} \right)\label{linanz1}, \quad m \in \mathbb{Z} \quad
\ee
and the order of the components is  $(\phi,r,t)$. The linearized equations of motion determine $g_{1,2}^m$ in terms\footnote{The solution for $g_{1,2}^m$ can be found in appendix \ref{solg12}.} of $g_3^m$, which itself obeys
\bea
&& \hspace{-2 cm} r^2 \p_r^2 g_3^m + 6 r \p_r g_3^m - \frac{r^z + \beta r + \a^2}{4 r^2} \, \p_t^2 g_3^m - \frac{i m}{2r}\, \p_t g_3^m + \frac{21 + 12 \nu - 9 \nu^2}{4}\, g_3^m = \non \\
&&  - \frac{9 \nu (\nu-1)}{8 r^2}\, (C_2^m(r) + C_1^m(r)\, t) - \frac{\nu}{8 r^4} \, C_3^m(r) + \frac{3 \nu}{4 r} \,( C_2'^m(r)  +  C_1'^m(r)\, t) \label{eqng3}
\eea
with
\be
C_1^m = \frac{2 i (C_3^m - r C_3'^m)}{3 m r} \;, \;\;\;\; m \neq 0.
\ee
Note that \eqref{eqng3} is a second order wave-equation for $g_3^m$ as expected, given that TMG has a single degree of freedom. So far,the functions $C_{2,3}^m(r)$ are arbitrary ``constants'' of integration, but it is also true that the choice \eqref{syncg} does not completely fix the gauge freedom. A thorough analysis of the gauge fixing is done in appendix \ref{gaugefixapp}. We can then write the most general solution to the above equations as
\be
h_{\mu\nu} = h_{\mu\nu}^{hom} + h_{\mu\nu}^p
\ee
where $h_{\mu\nu}^{hom}$ solves the homogenous equations of motion (where all the $C_i^m(r)$ have been set to zero), whereas $h_{\mu\nu}^p$ is the particular solution to the non-homogeneous equation. In view of their behavior with respect to time, the solutions to the homogenous equation will be called exponential or oscillating modes, whereas the particular solutions will be referred to as polynomial modes.

\subsection{Exponential modes}\label{monBH}

The simplest way to solve the homogenous equations of motion is by means of a Laplace transform
\be
h_{\mu\nu}^{hom} (\phi,r, t) = \int \frac{d\omega}{2\pi} e^{-i\om t} \tilde{h}_{\mu\nu}^{hom} (\phi,r,\om) \;, \;\;\;\;\;\; \om \in \mathbb{C}.
\ee
The solutions with $\omega \neq 0$ will be propagating since they cannot be gauged away.
%The solution for $g_{1,2}$ is given by \eqref{g12inapp} with the replacement $\p_t \r - i \om$.
The equation of motion for $g_3$ now reads
\be
r^2 g_3'' + 6 r g_3' + \frac{1}{4}\left( \omega^2\, r^{z-2} + 4(4+3z-z^2) + r^{-1} \omega (\beta \omega -2m) + r^{-2} \alpha^2 \omega^2 \right) g_3 =0 \label{eqg3}.
\ee
Solutions to the above equation for the $z=2$ case will be explored in section \ref{expsolns}.

\subsection{Polynomial Solutions}

In addition to the exponential modes, one can also obtain a set of solutions which are polynomial in time. A careful treatment of gauge-fixing for these modes can be found in appendix \ref{gaugefixapp}. We only retain particular solutions that blow up slower than the background in the limit $r \r \infty$; more precisely we require that $g_{1}^m \sim r^{a}$ with $a < 2$. In this case, the completely gauge-fixed solutions read
\begin{eqnarray}
g_1^m &=& \g^m r \, t + + r \beta^m \d^0_m + a^m \d^0_m - \frac{2 c_3^m \nu r t^2}{3 \nu-1}+ \frac{i m \nu c_3^m}{3 \nu -1} t +  \frac{c_3^m \nu \, r^{\frac{3 \nu-1}{2}}}{2 (3 \nu -1)} + \frac{c_3^m \a^2 \nu}{2 r (3 \nu -1)}, \non \\
g_2^m &=& \frac{\nu c_3^m t}{(3\nu-1) r} \;, \;\;\;\;\; g_3^m=0. \label{gamgauge}
\end{eqnarray}
The above notation signifies that the modes proportional to $e^{- i m \phi} (\b^m r + a^m)$ can be gauged away whenever $m \neq 0$.

The $c^m_3 r \,t^2$ and $\g^m r \, t $ modes are potential sources of instabilities due to their unbounded growth in time. To see whether they are indeed part of the physical spectrum we should first check finiteness of their norm, which is the subject of Part II.

\section*{{Part II: Physical Requirements}}

At this point, we would like to discuss the physical conditions to be imposed on the above modes. We require three properties:

\begin{itemize}
\item  physical excitations should have a finite symplectic norm. The norm of a linear mode is defined \`a la Lee-Wald \cite{Lee:1990nz} as
\be
\mathcal N \equiv \Omega_{\Sigma_t} ( h ,h) = -i \int_{\Sigma_t} \Omega[ h, h^* ; \bar g ], \label{symplnorme}
\ee
where $\Omega$ is given in the Appendix~\ref{symplApp} and $\Sigma_t$ is a constant $t$ hypersurface.
\item physical modes should have a finite flux at infinity. The symplectic flux at infinity is defined as
\be
\mathcal F \equiv \Omega_{\Sigma_r} ( h ,h) = -i \int_{\Sigma_r} \Omega[ h, h^* ; \bar g ], \label{symplflux}
\ee
where $\Sigma_r$ is a constant $r$ hypersurface that we send to infinity.  Self-adjointness of the boundary Hamiltonian requires that the flux at infinity be zero, which leads to the conservation of the symplectic norm. It is however interesting to also describe travelling wave solutions which have a non-zero flux at infinity.
\item
finally, a physical mode should be infalling at the horizon.
\end{itemize}

There are three ways that modes in the spectrum can represent classical instabilities. First, positive frequency modes can have negative norm (be ghosts) or modes can have negative energy with respect to the $\partial_t$ Killing vector. This would imply that our proposed state is unstable or is not the lowest energy configuration. This is what happens for linear fields in TMG around AdS$_3$ with the positive sign of Newton's constant \cite{Deser:2002iw,Carlip:2008jk,Grumiller:2008qz}. Second, the modes can acquire an unbounded growth in time, thus creating a large backreaction at late times. This kind of instability occurs for example around small Kerr-AdS black holes \cite{Press:1972aa,Cardoso:2004nk}. Finally, closed timelike curves might develop at late times starting from a regular initial data spacelike surface. This in turn leads to pathologies in the stress-tensor of quantum fields defined on the geometry \cite{Kay:1996hj}.

Our task is to analyze all solutions to the linearized equations of motion that we found in Part I and classify them according to their norm and their potential to induce various types of instabilities.

\subsection{Null warped oscillating gravitons}\label{expsolns}

For the null case $z = 2$, the exact solution to \eqref{eqg3} for $\alpha \neq 0$ reads
\be
g_3^m(r) = \frac{C_+}{r^2} {\mathcal{M}} \left(\frac{\beta \omega -2 m}{4 i \alpha  },\frac{\sqrt{1-\omega^2}}{2};\frac{\omega  i \alpha}{r}\right)+\frac{C_-}{r^2} {\mathcal{M}} \left(\frac{\beta \omega -2 m}{4  i \alpha },-\frac{\sqrt{1-\omega^2}}{2};\frac{\omega  i \alpha}{r}\right)\label{modeexp1}
\ee
where ${\mathcal{M}} (a,\pm b;x)$ are linearly independent Whittaker functions and $C_\pm \in \mathbb{R}$.
%\textbf{We should also pay attention to the special case $\omega = \pm 1$, $\beta \omega = 2 m$.}
% For $\beta \omega = 2 m$ and $\omega \neq \pm 1$, we find}
%\be
%g_3(r) =( \frac{\alpha\, \omega}{r})^{5/2} \left( C_- J_{\lambda_- } (\frac{\alpha\, \omega}{2 r}) \Gamma(1+\lambda_- )  +C_+ J_{\lambda_+ } (\frac{\alpha \, \omega}{2 r}) \Gamma(1+\lambda_+ )  ) \right). \label{modeexpbetamm}
%\ee
%with $\lambda_{\pm}= \pm \frac{\sqrt{1-\omega^2}}{2}$.
% ****What was in the file before ****
%\be
%g_3(r) = C_+ r^{-\frac{5}{2} - \frac{\sqrt{1-\omega^2}}{2} } + C_- r^{-\frac{5}{2} +  \frac{\sqrt{1-\omega^2}}{2} }.
%\ee
%****************************************
It is also worth noting that when $\omega = \pm 1$ and $\beta \omega = 2 m$, the solution becomes:
\be
g_3(r) =\left(\frac{\alpha}{r}\right)^{5/2} \left[ C_+ J_0 \left(\frac{\alpha}{2 r}\right) +  C_- Y_0 \left(\frac{\alpha}{2 r}\right) \right],
\ee
where $J_\nu$ and $Y_\nu$ are the Bessel functions of first and second kind respectively.

\subsubsection*{Asymptotic behavior near $r=\infty$}

As $r \r \infty$, the solution (\ref{modeexp1}) behaves as
\be
g_3(r) = C_+ (i \a \om )^{\k_+} r^{-\frac{5}{2} - \frac{\sqrt{1-\om^2}}{2}} \left(1+ \O(r^{-1}) \right) +  C_- (i \a \om )^{\k_-} r^{-\frac{5}{2} + \frac{\sqrt{1-\om^2}}{2}} \left(1+ \O(r^{-1}) \right) \label{eq:99}
\ee
where $\k_{\pm}= \half \,(1\pm \sqrt{1-\om^2})$. The main feature of this asymptotic behavior is its dependence on the frequency along the null direction. All $\om \in \mathbb{R}$ with  $\omega > 1$ lead to a complex exponent and thus the corresponding solutions are traveling waves. We choose the branch $Re\left(\sqrt{1-\omega^2}\right)\geq 0$. The solution multiplied by $C_+$ will subsequently be referred to as the fast-falling mode and the one proportional to $C_-$, as the slow-falling mode.

The symplectic flux  and norm of the fast/slow-falling mode around  \eqref{eq:98} behave asymptotically as
\begin{equation}
\mathcal F_{\pm} \sim |C_{\pm}|^2r^{\mp Re(\sqrt{1-\omega^2}) },\quad \mathcal N_{\pm} \sim |C_{\pm}|^2 \int dr \; r^{-1 \mp Re(\sqrt{1-\omega^2}) }.
\end{equation}
The contribution of the slow falling mode to the symplectic norm acquires an infinite contribution from the integration at large radii, and thus we must set $C_-=0$. The fast falling mode contributes a finite amount to the symplectic norm in the region $r\gg 1$, except when $\om \in \mathbb{R}$, $\om >1$. In this case, both fluxes $\mathcal{F}_{\pm}$ are finite but the norm is divergent. These modes represent traveling waves which carry away energy from the bulk to the boundary. For $\omega = 1$, only the fast-falling branch is allowed, since the metric and flux of the other branch exhibit  logarithmic divergences.

To summarize, the analysis of the asymptotic behavior of the perturbations near $r=\infty$ in asymptotically null warped AdS$_3$ allows only the fast falling modes - $ C_+$ - for all values of $\omega$ except $\om \in \mathbb{R}$ with  $\omega \geq 1$. Perturbations with real $\omega \geq 1$ represent traveling waves which have nonzero flux through  the boundary. All perturbations have a characteristic frequency-dependent falloff at large radius.

\subsubsection*{Near-horizon behavior}

The Whittaker functions admit an irregular singular point at $r = 0$. In the limit $r \rightarrow 0$, the fast falling mode  behaves as
\be
g_3^m \sim C_+\frac{\Gamma (2 b +1)}{(i \om \a)^a \Gamma(b-a+\half) } e^{\frac{i \alpha  \omega}{2 r} + \mathcal{O}(\log r)} +C_+ \frac{ \Gamma(2b+1)}{(i \om \a)^{-a}\Gamma (a+b+\half)}  e^{-\frac{i \alpha  \omega}{2 r} + \mathcal{O}(\log r)} +\ldots \label{solnh}
\ee
where we have used the shorthand notation
\be
a=\frac{\b \om -2 m}{4 i \a} \;, \;\;\;\;\; b = \frac{ \sqrt{1-\om^2}}{2}\;, \;\;\;\;\; \a >0.
\ee
Imposing that the modes be purely infalling at the horizon for $Re (\om) >0$ sets to zero the coefficient of the second exponential and thus leads to the following quantization condition on the frequency
\be
m = 2 \pi i\,T_R  \left[ n + \frac 1 2 \left(1+\sqrt{1-\omega^2} \right) \right] + \frac{\beta\omega}{2}, \quad n = 0,1,2,3,\ldots \label{quasin}
\ee
where $T_R$ is the right-moving temperature given in \eqref{temperatmax}.

Since we have chosen the fast-falling mode with $Re\left( \sqrt{1-\omega^2}\right) \geq 0$, it is not hard to see that all solutions to this equation must have $Im(\om) <0$. The mode is then exponentially growing at the horizon and therefore not normalizable. There are therefore no normalizable quasi-normal modes. Infalling modes blow up either at the horizon or at infinity.

It is nevertheless interesting to note that the above expression resembles a pole of the Green's function in a  conformal field theory \cite{Birmingham:2001pj} upon identifying:
\be
h_R = \frac{1}{2}+\frac{1}{2}\sqrt{1-\omega^2}, \quad  q_R  = \omega,\quad  \Omega_R = \frac{\beta}{2}, \quad  \omega_R = m
\ee
such that
\be
\omega_R = 2\pi i\, T_R  \left( n + h_R \right) + q_R \Omega_R, \quad n = 0,1,2,3,\ldots
\ee
The expression for the `conformal dimension' of the operator `dual' to the TMG massive graviton takes the familiar form encountered in non-relativistic AdS/CFT \cite{Son:2008ye} and Kerr/CFT \cite{Hartman:2009nz}. Notice that the structure suggests the existence of a chemical potential proportional to $\beta$, whose appearance we will further discuss in section \ref{SpWAdS}.

Similar relations have been shown to hold for scalar and vector perturbations of the null warped black hole backgrounds \cite{Chen:2009hg, Chen:2009cg} as well as more general considerations of extremal rotating black holes \cite{Chen:2010ni,Anninos:2010gh}.

\subsection{Null warped polynomial solutions}
\label{pertnull2}

Now let us turn our attention to the solutions which are polynomial in time for $z=2$. The particular solution with  eigenvalue $m \in \mathbb Z$ under $-i \p_\phi$ and which is subleading with respect to the background metric takes the form
\begin{eqnarray}
g_1^m(r,t) &=& \g^m \, r\, t + a^m \delta_{m,0} + r \beta^m \delta_{m,0} - c_3^m r t^2 + \frac{c_3^m \a^2}{4 r} + \frac{i c_3^m m}{2} \, t \\
g_2^m(r,t) &=& \frac{c_3^m t}{2 r}\;, \;\;\;\;\; g_3(r,t) =0.
\end{eqnarray}
The non-axisymmetric perturbations with $m \neq 0$ depend on two arbitrary constants $c_3^m,\g^m$ for each $m$. The axisymmetric ($m = 0$) perturbations depend on four arbitrary constants $c_3^0,\g^0,a^0,\beta^0$. None of these parameters can be removed by a diffeomorphism.

The symplectic norm of the perturbation $c_3^m$ turns out to be infinite because of a divergence at the horizon $r=0$. We therefore have to reject this mode. The other perturbations have vanishing norm and a vanishing symplectic product between themselves. The linear solutions with $\g^m \neq 0 $ are growing linearly in time and therefore represent marginal instabilities of the black hole geometries. The physical implications of the existence of this mode will be discussed in section \ref{ASG}.

\section*{Part III: Null z-warped modes}

Finally, we briefly discuss linearized solutions about the null z-warped background.

\subsection{Null z-warped oscillating gravitons}

The explicit solutions to \eqref{eqg3} for general $z$ have not been found. For the case $z > 2$ the asymptotic behavior of $g_3^m$ is
\be
g_3 \sim C_1 \, r^{-\frac{5}{2} - \frac{z-2}{4}}  \cos{\left(\frac{\,r^{\frac{z-2}{2}}\omega }{z-2}\right)}+C_2 \, r^{-\frac{5}{2} - \frac{(z-2)}{4}}  \sin{\left(\frac{\,r^{\frac{z-2}{2}}\omega }{z-2}\right)}
\ee
which is qualitatively very different from the null warped case in the sense that there is no $\omega$-dependence in the envelope function.

Using the Lee-Wald symplectic norm (see Appendix \ref{symplApp}) we find
\be
\mathcal N \sim r^{z-2}  ,\qquad \mathcal F \sim r^{\frac{z-2}{2}}
\ee
which implies that for $z>2$ both the symplectic norm and the symplectic flux  acquire a divergent contribution from infinity. Thus we are lead to the conclusion that all modes are non-normalizable\footnote{There remains, nevertheless, a possibility that there exist a more suitable norm than the Lee-Wald one. For example, for asymptotically anti-de Sitter spacetimes, the counterterms in the action may render the symplectic norm finite through the addition of a boundary term whereas the symplectic flux, which is always finite, is independent of counterterms \cite{Compere:2008us}. For asymptotically anisotropic spacetimes such as null z-warped geometries, we leave open the question whether boundary contributions might exist, which cancel the divergences by modifying the measure.} and we should discard them. Note that this behaviour is also very different from that of perturbations in asymptotically anti-de Sitter space (or in Schr\"{o}dinger spacetimes), where at least one of  the two linearly independent solutions to the wave equation is normalizable.

\subsection{Null z-warped polynomial solutions}

The particular linearized solution to the equations of motion for $z>2$ takes the form
\be
g_1^m=- \frac{2 c_3^m \nu r t^2}{3 \nu-1} + r \g^m \,t \,\delta_{m,0} + \frac{i m \nu c_3^m}{3 \nu -1} t + \frac{c_3^m \nu }{2 (3 \nu -1)} r^{\frac{3 \nu-1}{2}} + r \beta^m \delta_{m,0} + a^m + \frac{c_3^m \a^2 \nu}{2 r (3 \nu -1)},
\ee
\be
g_2^m = \frac{\nu c_3^m t}{(3\nu-1) r} \;, \;\;\;\;\; g_3^m=0.
\ee
Here we have chosen a different gauge from the one used in  \eqref{gamgauge}, which is accessible for $z>2$ only. The symplectic norm of the $c_3^m$ mode is infinite as before because of a divergence at the horizon. One thus has to reject this mode. The symplectic product of all remaining polynomial modes between themselves is vanishing, as in the $z=2$ case. As before, the presence of the $\g^0 r \, t $ mode is quite worrysome.

\section{Asymptotic analysis of null warped spacetimes}\label{ASG}

Having discussed the linearized modes around the black holes, we proceed by proposing a consistent set of boundary conditions for null warped
spacetimes ($z=2$) and the associated asymptotic symmetry algebra. Before beginning the analysis, we will comment on the physical content of TMG with
null warped spacetime asymptotics.

\subsection{Null warped excitations}

Part of the problem of finding the natural boundary conditions for null warped AdS$_3$ stems from the fact that a coordinate-independent definition of these spacetimes is not known and consequently, we do not know how to characterize the most general asymptotically null warped spacetime. Furthermore, there exists no analogue of the Fefferman-Graham theorem in these cases \cite{Skenderis:2009nt}, which in particular implies that we do not know the most general nonlinear solution to the asymptotic equations of motion.

We shall instead use our knowledge of the \emph{full} linearized spectrum to infer the properties of the nonlinear one. The main assumption that we will make is that modes that are normalizable/ non-normalizable in the linear theory will continue to be so in the nonlinear case, too.

A large portion of the linearized solution space consists of modes exponential (or oscillatory) in time, which are characterized by the frequency $\om \in \mathbb{C}$. We showed that none of the modes with $Im (\om) \neq 0$ are physical around the black hole solutions, since they are either not normalizable or they correspond to outgoing waves at the horizon. We will thus (boldly) assume that the $Im (\om) \neq 0$ modes are also discarded at the non-linear level.

Real-frequency solutions with $\om >1$ represent travelling waves which have a non-zero flux at infinity and infinite norm. They are therefore not present in boundary conditions with conserved asymptotic charges. The real frequency solutions with $\om \in (0,1)$ do not solve \eqref{quasin} so they are again excluded.

Note that modes with $\om \neq 0$ generically have falloffs that involve non-integer powers of $r$.
The non-exponential solutions only have falloffs in integer powers of $r$. Therefore, after having excluded all the exponential modes, we can make an Ansatz for the asymptotic form of the metric, which involves only integer powers of $r$. Subsequently, we solve the nonlinear asymptotic equations of motion order by order, finding that generic nonlinear non-normalizable solutions have $g_{\phi\phi} \propto r^2$, while normalizable ones have $g_{\phi\phi} \propto r$. The natural choice of boundary conditions that follows from this analysis is given by \eqref{bndnaive}.\footnote{One may object that we have rejected a large part of the spectrum  based on the supposed non-normalizability in the interior of nonlinear modes that we have not solved for. Instead, we should have used only non-normalizability due to a divergent contribution near infinity. Assuming that nonlinear exponential modes behave similarly near infinity as their linearized counterparts, such a condition immediately excludes all the slow-falling modes. This naturally leads to boundary conditions (roughly) looser that \eqref{bndnaive} by $r^{\half}$. But then one notices that the propagating modes with $\om \in (0,1)$ obey these boundary conditions but lead to infinite conserved charges. Then, we are lead again to \eqref{bndnaive}, this time only having used the asymptotic behavior of fields. Nevertheless, the only way to exclude the fast-falling modes with large $Im (\om)$ is by the assumption of divergent behavior in the interior.}
%\be
%g_{\phi\phi} = r^2 + \O(r) \;, \;\;\;\;\; g_{\phi t} = r + \O(1) \;, \;\;\;\;\; g_{rr} = \frac{1}{4r^2} + \O(r^{-3})\non
%\ee
%\be
%g_{r\phi} = \O(r^{-1}) \;,\; \;\;\;\; g_{rt}=\O(r^{-2})\;,\; \;\;\;\;g_{tt} = \O(r^{-1}).
%\ee

The boundary conditions \eqref{bndnaive} seem natural and are consistent, as we will soon see. Nevertheless, they suffer from the problem that they allow the existence of normalizable polynomial modes that grow linearly in time, parameterized by $\g$ below
\be
\frac{ds^2}{\ell^2} =  2 r d \phi dt + \left(  r^{2} + \beta  r + \gamma r
\, t + \alpha^2  \right) d \phi ^2  + \frac{dr^2}{4r^2}.   \label{cosm}
\ee
This is a solution of the full nonlinear TMG equations of motion; its linearized version around the black hole solutions has been found in section \ref{perts}. Starting from regular initial data on the spatial $t=0$ slice, this solution develops closed timelike curves at late or early times, depending on the sign of $\gamma$. Since the solution \eqref{cosm} is normalizable and carries finite energy, a priori we have no reason to exclude this mode. This in turn prevents us from having a consistent theory of topologically massive gravity in null warped AdS$_3$.

A way to render the theory (at least classically) consistent is to notice that the above are not the only consistent boundary conditions one can impose.\footnote{One may wonder why we are allowed to choose our boundary conditions; after all, if holography exists in null warped AdS$_3$, one should have a unique interpretation of the $\g r t$ mode. The multitude of boundary conditions are just a reflection of our ignorance about the correct initial value problem at spatial infinity in null warped AdS$_3$, and future work should clarify the meaning of each of them.} We will show that a consistent truncation of the boundary conditions exists where these solutions are  rejected.

To recapitulate, the following analysis is \emph{not} an attempt at a systematic analysis of the on-shell asymptotic
expansion of the metric. Instead, we will limit ourselves to finding a consistent
subsector
%\footnote{This subsector consists of metrics which can be  expanded  in integer powers of $r$ near the boundary. This  turns out to be a consistent truncation containing already many interesting features. Of course in general we should be allowing also the arbitrary powers of $r$ found in the linearized propagating solution, as well as log modes.}
of the most general asymptotic expansion which contains at least
the black holes and the solution \eqref{cosm}. For more details please consult appendix \ref{asyexp}.

\subsection{Boundary conditions and the ASG}

As was already noted, we start with the following set of boundary conditions
\be
g_{\phi\phi} = r^2 + \O(r) \;, \;\;\;\;\; g_{\phi t} = r + \O(1) \;, \;\;\;\;\; g_{rr} = \frac{1}{4r^2} + \O(r^{-3})\non
\ee
\be
g_{r\phi} = \O(r^{-1}) \;,\; \;\;\;\; g_{rt}=\O(r^{-2})\;,\; \;\;\;\;g_{tt} = \O(r^{-1})
\label{bndnaive} \ee
defining a phase space $\cal F$. We have fixed $\ell=1$ for convenience. All linearized propagating modes with $\om \in \mathbb{R}$ are excluded by these boundary conditions. On the other hand, the mode \eqref{cosm} that grows linearly in time is allowed. Furthermore, the phase space includes all the known black hole solutions.

The mode growing linearly in time is problematic because it produces closed timelike curves at either early or late times. We therefore propose an alternative set of boundary conditions which exclude it:
\be
g_{\phi\phi} = r^2 + \b r+ \O(1) \;, \;\;\;\;\; g_{\phi t} = r + \O(1) \;, \;\;\;\;\; g_{rr} = \frac{1}{4r^2} + \O(r^{-3})\non
\ee
\be
g_{r\phi} = \O(r^{-1}) \;,\; \;\;\;\; g_{rt}=\O(r^{-2})\;,\; \;\;\;\;g_{tt} = \O(r^{-1}).
\label{bndbeta} \ee
The black hole solutions are still allowed, but with a difference in interpretation: while
the boundary conditions \eqref{bndnaive} imply that the whole two-parameter family of black holes is part of the same phase space, in the new boundary conditions  \eqref{bndbeta}, the parameter $\b$ is part of the asymptotic definition of the spacetime and only the parameter $\a$ characterizes the black holes. This difference  is important in view of the fact (to be established in the sequel) that $\b$ does not enter the asymptotic charges of the black holes, so according to \eqref{bndnaive} it would be interpreted as hair. %See section \ref{nwcgq} for further comments.

\subsubsection*{Asymptotic symmetries}

Asymptotic symmetries are those diffeomorphisms that leave the asymptotic form of the metric invariant and under which the allowed field configurations carry nontrivial charges. The asymptotic Killing vectors of \eqref{bndnaive} are given by
\begin{equation}\label{chiass}
\xi[L,N] = -r L^\prime(\phi)  \,\p_r + L(\phi)\, \p_\phi + N(\phi) \, \p_t  \,  + \xi_{triv},
\end{equation}
where $L(\phi)$ and $N(\phi)$ are arbitrary functions of $\phi$ and primes now denote derivatives with respect to $\phi$. The trivial asymptotic vectors $\xi_{triv}$ take the form
\be \xi_{triv}= \mathcal{O}({1}) \, \p_r +  \mathcal{O}\left({1\over r^2}\right) \, \p_\phi + \mathcal{O}\left({1\over r} \right) \label{Triv}
\p_t. \ee
Subsequently, we can define the generators
\be \label{defgen}
L_n := \xi [  e^{i n \phi} , 0\, ] \;, \;\;\;\;\;
N_n := \xi[0,  e^{i n \phi}]
\ee
which obey the following Lie bracket algebra
\be
i [L_n,L_m] = (n-m) L_{n+m},
\quad i [L_n,N_m] = -m\, N_{n+m}, \quad [N_m,N_n] = 0.
\label{ASA}\ee
This is a semi-direct sum of a Virasoro algebra and a $\widehat{u(1)}$ current algebra, which form the proposed asymptotic symmetry group (ASG) of null warped AdS$_3$.

For the more restrictive boundary conditions \eqref{bndbeta}, the asymptotic symmetries are
\be
\xi^{(\b)} [L,N] = L(\phi) \p_{\phi} + N(\phi) \p_t - \left(r L'(\phi) + \frac{\b}{2}\, L'(\phi)+ N'(\phi)\right)\p_r \label{xibeta}
\ee
so the only difference from \eqref{chiass} consists of fixing the form of the leading trivial diffeomorphism. The Lie bracket algebra of the Fourier modes of $\xi^{(\b)}$ is consequently the same as \eqref{ASA}. As we will show in the sequel, all backgrounds allowed by the more restrictive boundary conditions \eqref{bndbeta} have zero $N_n$ charges, so in fact the part of the ASG parameterized by $N(\phi)$ (the current algebra) is trivial from the point of view of \eqref{bndbeta}.

\subsubsection*{Asymptotic charges}

The charges  ${\cal{Q}}_{N}$ and ${\cal{Q}}_{L}$ associated with the asymptotic Killing vectors \re{chiass} of the metrics belonging to the phase space ${\cal{F}}$ defined by the boundary conditions \re{bndnaive} can be computed using the formalism of \cite{Barnich:2001jy,Barnich:2007bf}. We refer to \cite{Compere:2008cv, Compere:2009zj} for the explicit formulae (see also \cite{Bouchareb:2007yx}).

In our case, the expressions for the charges are quite involved. A considerable simplification occurs when one works in the radial gauge
\be
g_{t r} = g_{\phi r} =0 \;, \;\;\;\;\; g_{rr} = \frac{1}{4r^2}
\ee
which can be always reached using an appropriate trivial diffeomorphism of the form \eqref{Triv}. The most general (on-shell) asymptotic form of the remaining components of the metric that only involves integer powers of $r$ is
\be
g_{\phi\phi}= r^2 + \left(-t^2 c(\phi)+  t \g(\phi)+ \beta(\phi)   \right) r + \left(-\frac{t^2}{2} \, c(\phi)^2 + \frac{t}{2} \, c(\phi) \g(\phi)+ \frac{t}{2} \, \partial_\phi\, c(\phi)+ a(\phi)\right) + \ldots \non
\ee

\begin{equation}
g_{\phi t}= r + {\gamma(\phi) \over 2} + {1 \over 16} \gamma(\phi)^2 r^{-1}+..., \quad g_{tt}=0.
\end{equation}
This asymptotic form defines the arbitrary functions $c(\phi)$, $\g(\phi)$, $\b(\phi)$ and $a(\phi)$. They are ``constants of integration'' that appear when solving the full nonlinear asymptotic equations of motion. Note that the dependence on the various coordinates is precisely the same as we found in the linearized analysis in section \ref{perts}, which gives us a further consistency check.
%The expressions for the remaining asymptotic metric components can be found in appendix \ref{charges}.

The charges are then given by
\begin{eqnarray}
{\cal{Q}}_N &=& - {1 \over 24 \, \pi \, G} \int_0^{2 \pi} d\phi N(\phi) \, c(\phi),
\\
{\cal{Q}}_L &=& { 1 \over 24\, \pi \, G} \int_0^{2 \pi} d \phi [ L^\prime(\phi) \,  \g(\phi)+\frac{1}{8} L(\phi) \, h(\phi)] ,
\end{eqnarray}
with
\be
h(\phi) \equiv  \g(\phi)^2 + 3\, c(\phi)^2 + 32\, a(\phi) - 12 \,c(\phi) \,\beta(\phi)
\ee
and they can be further decomposed into their Fourier modes via \eqref{defgen}.

\subsubsection*{Consistency and the asymptotic algebra}

We have verified that \re{bndnaive} form a consistent set of boundary conditions for the asymptotic symmetry algebra \re{ASA}. More precisely, these Virasoro and current symmetries yield finite, integrable and asymptotically conserved charges. Moreover, the boundary conditions \re{bndnaive} are preserved by this asymptotic symmetry algebra and the charges associated to trivial asymptotic Killing vectors are zero.

The charges $\mathcal{Q}_{L_n}$ generate through their Poisson brackets a centrally extended Virasoro algebra whose central charge is given by
\begin{equation}
c_R = {2 \ell}/{G}.  \label{cc}
\end{equation}
The charges $\mathcal{Q}_{N_n}$ generate through their Poisson bracket a current algebra with level $k$ zero
\begin{equation}
k =0.
\end{equation}
Since \eqref{bndbeta} are a subset of \eqref{bndnaive}, finiteness, conservation and integrability of the charges for the second proposed ASG follows from the above. Note, nevertheless, that in this case the asymptotic symmetry group only contains the Virasoro algebra with central extension \eqref{cc}; the current algebra has become pure gauge, since the boundary conditions \eqref{bndbeta} do not allow for any metric which is charged under $\mathcal{Q}_N$.

\subsection{Consistent truncation of the boundary conditions}

A consistent classical theory of asymptotically null warped spacetimes should rely both on consistent boundary conditions and on a well-posed initial value problem. In particular, smooth initial data on a spacelike slice obeying the boundary conditions should evolve into a metric obeying the conditions at all times. The question that we would like to now address  is whether the restricted boundary conditions \eqref{bndbeta} are dynamically consistent. That is, if we impose \eqref{bndbeta} at $t=0$ and appropriately fine-tune the  values of the metric and its derivative on the initial spacelike slice, does it follow that  these boundary conditions will be obeyed at later times?

A necessary requirement for this to happen is that the modes that obey \eqref{bndnaive} but violate \eqref{bndbeta} decouple from the rest. To show this, we first remind the reader the full normalizable spectrum of TMG with boundary conditions \eqref{bndnaive}:

\bi
\item the linear mode in time $\widehat{\g^m}$, described by  $\d g_{\mu\nu} =\d^{\phi}_{\mu}\,\d^{\phi}_{\nu}\; \g^m\, e^{i m \phi} \, r\, t  \equiv \widehat{\g^m_{\mu\nu}} $
\item the scalar hair $\widehat{\beta^0}$, described by $\d g_{\mu\nu} =\d^{\phi}_{\mu}\d^{\phi}_{\nu} \beta^0 r\equiv \widehat{\beta^0_{\mu\nu}} $
\item the mass of the black hole $\widehat{a^0}  $ described by $ \d g_{\mu\nu} = \d^{\phi}_{\mu}\d^{\phi}_{\nu}\; i\, a^0 \equiv \widehat{a^0_{\mu\nu}}$
\item the Virasoro modes $\d g_{\mu\nu}  = \mathcal L_{L_m} \bar g_{\mu\nu} \equiv \widehat{(L_m \bar{g})}_{\mu\nu}$ which belong to the asymptotic symmetry group.
%\footnote{We choose the Virasoro modes to be localized at the boundary, e.g. $L_m = (1-e^{-r})(e^{i m \phi} \partial_\phi - i m e^{i m \phi} r \partial_r)$.}
\ei

All these linear modes are tangent to the phase space of regular solutions in the sense that they possess a regular initial spacelike slice (i.e. a spacelike slice where both the metric and the extrinsic curvature are smooth). The non-zero symplectic inner products
- denoted by $\langle \, | \, \rangle$ - of these modes around the solution \eqref{cosm} are:\footnote{We sincerely hope that the freedom we have taken to take off the hats when their presence would have rendered the notation too cumbersome will not confuse or inconvenience our reader. }
\be
\langle L_m | \,\gamma^n \rangle = -\frac{c_R}{24}\, m \,\g^m \, \delta_{m,n}\;, \;\;\;\;\; \langle L_m |\, a^0 \rangle = -\frac{c_R}{6}\, a^0 \,\delta_{m,0}\;, \;\;\;\;\; \langle L_m |\, L_n \rangle = -\frac{c_R}{12}\, m\,(m^2+4\alpha^2 )\, \delta_{m,n}.\label{sympprod}
\ee
Here $c_R$ is given by \eqref{cc}. Note that the mode $\b^0$ has zero norm and zero inner product with everything else, similar to the behavior of a pure gauge mode. The mode $\g^m$ also has zero norm, but the fact that it has a nonzero symplectic product with the Virasoros means that our basis is not diagonal. Around the black holes with $\alpha^2 \geq 0$ this can be easily fixed, by defining
\be
\widehat{\gamma^{\perp,m}} \equiv  \widehat{\gamma^m}  + \frac{c_R \; m}{24 \langle L_m | L_m \rangle }  \,\widehat{(L_m \bar{g})} \quad \forall \;\; m \neq 0, \quad \widehat{\gamma^{\perp,0}} \equiv \widehat{ \gamma^0}.
\ee
The new mode has
\be
\langle \gamma^{\perp,m} | \, \gamma^{\perp,n} \rangle = \frac{c_R\; m}{48 (m^2 + 4\alpha^2)}\, |\g^m|^2 \,\delta_{m,n} \;, \;\;\;\;\; \langle \gamma^{\perp,m} |\, L_n \rangle =0.
\ee
The norm of $\gamma^{\perp,m}$ is positive for $m \geq 1$.  The symplectic norm of the Virasoro perturbations $|L_m\rangle $ is proportional to the central charge and always positive for $m \leq -1$ as it should.
%\bigskip

The purpose of the above analysis was to show that the problematic $\g^m$ modes (or, rather $\g^{\perp, m}$) can be decoupled from the rest, at least at the classical level. So can the somewhat more innocuous ``hair'' modes $\b^0$. Therefore, at the classical level, we can consistently set these modes to zero by adopting the  restricted boundary conditions \eqref{bndbeta} where $\beta$ is fixed.

These restricted boundary conditions are also compatible with the initial value problem. It is not a priori obvious that a mode can consistently be truncated this way. In the case at hand however, we do know all non-linear solutions corresponding to the linear modes not excluded by \eqref{bndbeta}. They are the black hole mass difference $a^0$ and the nonlinear version of the Virasoro modes, obtained by acting with the finite diffeomorphism associated with $L_m$. None of these modes depends explicitly on time. Then it is easy to show that if the initial data (metric and its first derivative) is compatible with the boundary conditions \eqref{bndbeta} from which the mode $\g^m$ is absent, then it would not appear at later times.

The conclusion of this classical analysis is that the boundary conditions \eqref{bndbeta} with fixed $\b$ are consistent dynamically and contain only null warped BH as regular solutions. This is similar to what happens in spacelike warped AdS$_3$ \cite{Anninos:2009zi}, where consistent boundary conditions also only allow for the existence of black holes. The main difference between the two theories is that in null warped AdS$_3$ the allowed spectrum is chiral, whereas in spacelike warped it is not, because in that case the current algebra in the ASG has a non-zero central charge.

% \subsection{Lack of a stable ground state : should be removed, but here are some more comments}

% A CFT should admit a stable $SL(2,\mathbb R)\times SL(2,R)$ invariant ground state. It is interesting to note that for $\alpha^2 = -1/4$, the Virasoro symplectic products \eqref{sympprod} take the usual form which suggests that the solution \eqref{metric} with $\alpha^2 = -1/4$ might be the ground state of the CFT. Further indication comes from the fact that the energy of that state is $-c_R/24$ as deduced from \eqref{Qphiz}. The solution contains however closed time-like curves and morevoer the unstable mode $\gamma^m$ for $m=\pm 1$ do not decouple from the Virasoro which signals an instability. The lack of a potential ground state other than the aforementioned unstable solution is puzzeling and worrying.

\section{Cardy's formula and chirality}\label{CardySec}

The existence of a centrally extended Virasoro algebra in the asymptotic symmetries is intriguing. In particular, it suggests that if a consistent quantum mechanical description of topologically massive gravity exists about the null warped backgrounds, it would be holographically equivalent to a (chiral half of a) two-dimensional conformal field theory.

In this section, we bring further support to this conjecture by showing that the entropy of null warped black holes can be reproduced from a chiral Cardy formula, where the central charge and the energy are read off from the spacetime asymptotics. We also gather together the evidence for chirality of the theory defined with boundary conditions \eqref{bndnaive} or \eqref{bndbeta}.

\subsection{Entropy match}

One of the generic features of such a CFT$_2$, assuming modular invariance of the partition function, is that the spectrum of states at large energy has a degeneracy given by  Cardy's formula:

\be
S_{CFT} = 2 \pi \sqrt{\frac{c_L E_L}{6}} + 2 \pi \sqrt{\frac{c_R E_R}{6}}. \label{cardy}
\ee
In our case, the real numbers $c_{L,R}$ are the (semi)-classical central charges of the Virasoros as read from the central extension of the ASG, and $E_{L,R}$ are the appropriate conserved charges. If the black hole is extremal (such as the case at hand), then either the left or the right-movers are in their ground state and only one term survives in \eqref{cardy}.

Thus, using the central extension of the Virasoro found in section \ref{ASG}, together with the conserved charge \eqref{adtcharge}
%\be c_R = \frac{2 \ell}{G} \;, \;\;\;\;\; \; E_R = \frac{\a^2 \ell}{3G} \ee
as basic inputs of the Cardy formula \eqref{cardy}, we observe that they precisely match the macroscopic entropy of the null warped black holes, i.e.
\be
S_{BH} = \frac{2 \pi \a \ell}{3 G} = S_{CFT}.
\ee
We take this as evidence that the putative holographic dual to TMG in null warped AdS$_3$ has a consistent chiral subsector which is acted upon by a Virasoro algebra. %In any case, this matching points towards the fact that the proposed boundary conditions and ASG are correct.

\subsubsection*{Black hole hair}

A point that may be interesting to discuss concerns the difference between the boundary conditions \eqref{bndnaive} and \eqref{bndbeta} in the context of black hole degeneracy.

From the point of view of \eqref{bndnaive} the parameter $\b$ that appears in the black hole metric \eqref{metric} represents classical hair, because it does not appear in any of the conserved charges. Thus, for each value of the conserved charge $E_R$ , we have a continuous infinity (classically)  of black hole solutions, parameterized by $\beta \geq 0$. Quantum mechanically, we expect the values of the hair $\b$ to be quantized, but they will still contribute (most likely logarithmically) to the entropy. Their presence would not affect the leading entropy given by Cardy's formula, but if one had a precise definition of the quantum theory that would make subleading contributions computable, their contribution to the partition function would have to be taken into account. Moreover, one would have to find on the field theory side modes that would correspond to the spacetime hair.

On the other hand, from the point of view of \eqref{bndbeta}, the parameter $\beta$ is part of the boundary data, so for each value of the charge only one black hole solution exists. Thus, whether hair exists or not depends on the interpretation of the boundary conditions \eqref{bndnaive} or \eqref{bndbeta} in a proper asymptotic analysis. This is an interesting question that we leave for the future.\footnote{It is amusing to point out that hair also exists in chiral gravity. Namely, all BTZ black holes, parameterized by two constants $M$ and $J$ (the values of mass and angular momentum as computed in Einstein gravity), are solutions of TMG at the chiral point, with charges $L_0 = 0 \,, \;\bar{L}_0 = M+J$. For every fixed $M+J$ there exists (classically) an infinity of solutions $(M+n,J-n)$ with the same conserved charges  which are not trivially diffeomorphic to the original solution. It would be interesting to see how this observation fits in with the analysis in \cite{Maloney:2007ud,Maloney:2009ck}.}

\subsection{Relation to spacelike warped black holes}\label{SpWAdS}

Another piece of evidence that TMG in the null warped AdS$_3$ background may be dual to a chiral half of a CFT$_2$ comes from viewing null warped AdS$_3$ as a limit of spacelike warped AdS$_3$.  The spacelike warped solutions exist for $\nu >1$ and TMG in such spacetimes has been conjectured to be holographically dual to a CFT$_2$ with left/right-moving central charges given by \cite{Anninos:2008fx}
\be
c_L = \frac{4 \nu }{(\nu^2+3) }\frac{\ell}{G}, \quad c_R = \frac{(5\nu^2+3)}{(\nu^2+3)} \frac{\ell}{G}.
\ee
Black holes in these spacetimes are conjecturally dual to thermal states in the above CFT$_2$ with certain left/right-moving temperatures $T_{L,R}$.

As mentioned in \cite{Anninos:2008fx}, null warped AdS$_3$ can be obtained as a pp-wave limit of the other warped AdS$_3$ spacetimes. Below we show how to obtain the null warped black holes as a  limit of the spacelike warped black holes. The latter are described by the metric\footnote{We have changed coordinates with respect to \cite{Anninos:2008fx}, by letting $r = \bar{\rho} + \frac{r_++r_-}{2} + \frac{(r_+-r_-)^2}{16\bar\rho}$.}
\be
ds^2 = d\bar{t}^2+\frac{ d\bar{\rho}^2}{(\nu^2+3)\bar{\rho}^2} + 2\nu\left(\bar{\rho} + \tau_L + \frac{\tau_R^2}{4 \bar{\rho}}\right)d\bar{t}\, d\phi + g_{\phi\phi} (\bar{\rho}) \, d \phi^2
\ee
with $\phi \sim \phi + 2 \pi$ and
\be
g_{\phi\phi} = \frac{3}{4} \, (\nu^2-1) \bar{\rho}^2 + 2 \nu^2  \tau_L\, \bar{ \rho} + \tau_L^2 \nu^2 + \frac{5 \nu^2+3}{8} \tau_R^2 + \frac{\tau_R^2\, \tau_L \nu^2}{2 \bar{\rho}} + \frac{3\tau_R^4(\nu^2-1)}{64 \bar{\rho}^2}
\ee
where the two parameters $\tau_L$ and $\tau_R$ are proportional to the left and right-moving temperatures in the proposed dual CFT$_2$
\be
\tau_L = \frac{4 \pi \ell\, T_L}{\nu^2 + 3} \;, \;\;\;\;\;\tau_R = \frac{4 \pi \ell\, T_R}{\nu^2 + 3}.
\ee
Defining the new coordinates $t$ and $\rho$ by
\be
\rho = \l \, \bar{\rho}\;, \;\;\;\;\; t = \frac{\nu \, \bar{t}}{\l} \;, \;\;\;\;\; \l = \frac{ \sqrt{3 (\nu^2-1)}}{2} \label{resc}
\ee
and taking the limit $\nu \r 1^+$ and $\tau_L \r 0$ while keeping
\be
\b \equiv \frac{2 \tau_L}{\l} \;, \;\;\;\;\; \a \equiv \tau_R
\ee
fixed, results in the null warped black hole solutions \eqref{NullBH}.  According to the conjecture in \cite{Anninos:2008fx}, the null warped black holes correspond to thermal states with
\be
T_L =0, \quad T_R =  \frac{\a}{\pi \ell}
\ee
in a CFT$_2$ with central charges
\be
c_L = \frac{\ell}{G}\;,  \quad c_R = \frac{2\ell}{G}.
\ee
Note that $T_{L}$, $T_{R}$ and $c_R$ precisely match those in the previous analysis.

One can also compare the spectrum of quasinormal modes of the null warped and spacelike warped black holes. The latter have been worked out (for scalars) in \cite{Chen:2009cg} and read
\be
\om_L = -2 \pi i T_L (n_L + h_L), \quad  \om_R =  - 2 \pi i T_R (n_R + h_R) + q_R \Omega_R, \quad n_L, n_R \in \mathbb{Z}^+ \label{omlr}
\ee
where the various quantities are
\be
\om_R \equiv - \frac{m}{\ell}\;,  \quad \om_L \equiv \frac{\nu^2}{2\ell} \, \b \om, \quad  q_R \Omega_R \equiv  - \frac{\nu^2}{2\ell} \, \b \om %\frac{\nu \tau_L \bar{\om}}{\ell} =  \frac{\nu^2}{2\ell} \, \b \om
\ee
and
\be
h_L = h_R = \half + \half \, \sqrt{1- \frac{12 (\nu^2-1)}{(\nu^2+3)^2}\, \bar{\om}^2 + \frac{4 \ell^2 M^2}{\nu^2+3}}. \label{hgrav}
\ee
In \cite{Anninos:2009zi}, the conformal dimension of the operator dual to the graviton in spacelike warped AdS$_3$ was shown to be given by \eqref{hgrav} with $M^2 \ell^2 = -3 (\nu^2-1)$. In the limit $\nu \r 1^+$, \eqref{omlr} becomes precisely\footnote{Note that we are matching the equation for the quasinormal modes of the null warped \emph{graviton} to the $\nu \r 1$ limit of the quasinormal modes for a \emph{scalar} of the expected mass in spacelike warped AdS$_3$. That the TMG massive graviton resembles in certain ways  a scalar has already been noted in \cite{Deser:1981wh,Anninos:2009jt}.} \eqref{quasin}. We therefore see that the origin of $\b$ resides in the left-moving angular potential of the spacelike warped black holes.

Noting that the graviton in TMG around spacelike warped resembles a scalar field of mass $M^2 \ell^2 = -3 (\nu^2-1)$ \cite{Kim:2009xx,Anninos:2009zi}, the quasi-normal modes of the graviton take the same form as above.

Now let us take a look at the asymptotic generators in spacelike warped AdS$_3$, which consist of a right-moving Virasoro algebra and a current algebra \cite{Compere:2009zj,Compere:2008cv}
\be
L_n = e^{i n \phi} (\p_\phi - i n \bar{\rho} \, \p_{\bar{\rho}} ) \;, \;\;\;\;\; \bar{N}_n = e^{i n \phi} \p_{\bar{t}}
\ee
Under the rescaling \eqref{resc}, the Virasoro generators do not change. Nevertheless, the would-be current algebra in null warped AdS$_3$ is given by:
\be
N_n =  e^{i n \phi} \p_{t} = \l \bar{N}_n
\ee
and thus all generators vanish in the limit $\l \rightarrow 0$. This is in agreement with our results that in null warped AdS$_3$ the current algebra is trivial.

\subsection{Remarks on chirality}\label{nwcgq}

The classical analysis that we have performed so far, of both linearized perturbations and the asymptotic symmetry algebra, indicates that the classical theory of TMG in null warped AdS$_3$ with boundary conditions \eqref{bndnaive} or \eqref{bndbeta} is \emph{chiral}. We have three pieces of supporting evidence:
\begin{enumerate}[(i)]
\item  the (known) black hole spectrum is chiral
\item all normalizable modes are charged only under the (chiral) Virasoro algebra
\item the current algebra acts trivially
\end{enumerate}
In the above, `chiral' means `only charged under  the Virasoro symmetries'.

\subsubsection*{(i) The black hole spectrum and a Birkhoff-like theorem}

All known black hole solutions to TMG at $ \mu \ell = 3$ with null warped asymptotics are given by \eqref{metric}. They are characterized by only one conserved charge, which is the zero-mode of the chiral Virasoro that forms the asymptotic symmetry group.

It is interesting to ask whether other black hole solutions may exist with the same asymptotics. In appendix \ref{birkhoff} we manipulate the formalism developed in \cite{Bouchareb:2007yx} to prove that any analytic solution of TMG with $\mu \ell = 3$ that admits two commuting Killing vectors and has the same asymptotics as the black holes must be one of the null warped black holes.

\subsubsection*{(ii) Chirality of the normalizable spectrum  }

In the previous sections,  we analyzed the dynamics of the theory from two perspectives: at the linear level around the black hole geometries and at the non-linear level via an asymptotic expansion (restricted to integer powers in the radial fall-off). Exploiting these two cases, we can infer the full \emph{classical} spectrum of the theory. Below is a review of our  conclusions.

The linear theory around the black holes contains two types of modes: exponential modes and polynomial ones. The boundary conditions \eqref{bndnaive} exclude all slowly-falling exponential modes and all the travelling waves. The boundary conditions alone do not, nevertheless, exclude the fast-falling modes. As we have seen in section \ref{perts}, these modes are not normalizable because of a divergence at the horizon. Therefore, we expect that any nonlinear solution that has noninteger powers of $r$ turned on or exhibits exponential behavior in time will be singular. We will reject all such modes under this assumption.

The polynomial mode labeled by $c_3^m$ in the linearized theory and by $c(\phi)$ in the asymptotic nonlinear solution (and proportional to $t^2$) is also not normalizable due to a divergence near the horizon. Therefore it is also excluded from the spectrum of normalizable solutions. Thus, all normalizable modes allowed by the boundary conditions have vanishing current charges\footnote{Although the current algebra generators depend only on the modes of the coordinate $\phi$, just like the Virasoros, it is not clear whether they belong to the same chirality subsector of the dual theory, as we will discuss shortly. } $\mathcal{Q}_N$, and consequently the spectrum of propagating modes is chiral.

\subsubsection*{(iii) Triviality of the current algebra}

The only modes whose chirality we still have to check are the pure large gauge modes. Namely, if we adopt \eqref{bndnaive} then the ASG contains a current algebra with level zero. The only backgrounds that could be charged under the current algebra are those diffeomorphic to the $c^m_3$ solution, which suffer from divergences in the interior. Therefore, no normalizable mode in the theory is charged under the $N_n$, so the large diffeomorphisms $N_n$ are pure gauge (they act trivially on the phase space) and are not contained in the asymptotic symmetry group.

That the current algebra should be trivial is corroborated by our argument regarding null warped black holes as a particular $\nu \r 1$ limit of spacelike warped black holes.

\subsubsection*{Discussion}

The above arguments seem to strongly suggest that TMG in null warped AdS$_3$ is chiral. As usual, this statement should contain a clear specification of the boundary conditions, which in our case are \eqref{bndbeta}. The careful analysis of perturbations and asymptotic behavior that we have performed so far supports this claim. Moreover, the ASG that we found supports the idea that TMG in null warped AdS$_3$ with appropriate boundary conditions is dual to a chiral half of a CFT$_2$, which is most possibly the $\nu \r 1$ limit of the CFT$_2$ conjectured to be dual to the spacelike warped black holes.

Nevertheless, it is probable that the boundary conditions \eqref{bndnaive} or \eqref{bndbeta} are not the ones that capture the most interesting physics in the null warped spacetime. Firstly, they leave out the fast-falling exponential modes, which would have been the interesting propagating modes of the theory. Secondly, the naive limit from spacelike warped AdS$_3$  seems to indicate that TMG in null warped AdS$_3$
should be dual to a CFT$_2$ with $c_R = 2 c_L = 2 \ell G^{-1}$, but no vestiges of the left-movers were seen in our analysis, except for the mysterious hair $\b$. It is not unlikely that  non-standard boundary conditions might exist, which would lead to an asymptotic symmetry group that contains both the left- and the right-movers. Note that if this is the case, then it doesn't make much sense to talk about a chiral TMG theory in null warped AdS$_3$, since we would have unnaturally set the left movers to their ground state. Rather than speculating any further, we leave the resolution of this issue to future work.

\bigskip

To summarize, we have presented various arguments that show that TMG  in null warped AdS$_3$ may be chiral at the classical level. Such a theory would have no propagating degrees of freedom, given that the massive graviton of TMG is excluded by the boundary conditions \eqref{bndbeta} and the remaining time-dependent modes decouple. This situation is reminiscent of chiral gravity, where the Brown-Henneaux boundary conditions disallow the propagating logarithmic mode, and the remaining spectrum is again chiral. We leave open the question of whether the chirality of the spectrum is preserved in presence of quantum corrections.
Our analysis is not very illuminating in that sense, and the clever argument that was applied in \cite{Strominger:2008dp,Maloney:2009ck} to decouple chiral gravity from the remaining modes of log gravity cannot be applied here, for the simple reason that the ASG analysis is not appropriate to uncover holography in spacetimes where flux can pass through the boundary at infinity.

\section{Asymptotic analysis of the null z-warped spacetimes}\label{ASGz}

In this section, we make a proposal for the asymptotic symmetry group of z-warped spacetimes. We  check that this proposal leads to finite, integrable and conserved charges in the very restricted phase space that consists of black hole solutions and large diffeomorphisms thereof. This construction is enough to provide a representation of the asymptotic Virasoro algebra, to derive its central charge and to match the entropy of these black holes using Cardy's formula. The problem of finding consistent boundary conditions for asymptotically z-warped spacetimes will not be settled here due to its technical difficulty.

For purely orientational purposes,  a tentative definition of the boundary conditions is presented below: %{\bf (Proposal for $g_{tt}$ required for Birkoff theorem)}
\be
g_{\phi\phi}= \O(r^z) \;, \;\;\;\;\; g_{\phi t} = r + \O(1) \;, \;\;\;\;\;g_{\phi r } = \O(r^{-1}),\non \ee
\be
g_{rr}=\frac{1}{4r^2}+O(r^{-3}), \qquad g_{tt}=O(r^{-1}).\label{bctent}
\ee
Note that such boundary conditions allow the unstable solutions
\be
\frac{ds^2}{\ell^2} =  2 r d \phi dt + \left(  r^{z} + \beta r + \gamma \, r
\, t + \alpha(\phi)^2  \right) d \phi ^2  + \frac{dr^2}{4r^2}\label{eq:cosmz}
\ee
which generalize to any $z$ the solution \eqref{cosm}. These backgrounds turn out to have finite, integrable and conserved Virasoro charges. The form of the ``tentative'' boundary conditions is partly inspired by the boundary conditions in \cite{Guica:2008mu}, partly by matching to the symmetry
generators  in the $z=2$ case (\ref{chiass}),\footnote{Note that this set of generators is not the one which is found by solving Killing equations on the z-warped background at first order in the radial expansion along the procedure outlined in \cite{Barnich:2006av,Compere:2007az} because here the leading metric coefficient $g_{\phi\phi}$ is allowed to change.} and mostly by the fact that they give the correct central charge that matches Cardy's formula.\footnote{The boundary conditions where $g_{\phi\phi} = r^z + \O(r)$, $g_{t\phi}=\O(r)$ also have a Virasoro as ASG. Its central extension, however, yields a central charge which does not match, via Cardy's formula, the black hole entropy.} The reader is thus entitled to object that we are making an ``educated guess''. Future work should corroborate the correctness of this guess.

\subsection{Black hole phase space}

The phase space we will consider is given by acting on the axisymmetric black holes \re{NullBH} with the finite diffeomorphisms generated by the asymptotic Killing vectors $\xi[L,0]$ defined in \eqref{chiass}. The resulting metric is just the black hole metric \re{NullBH} with the following coefficients differing
\be
g_{r\phi} = - \frac{\mathfrak{L}''}{4\mathfrak{L}' r}\;, \;\;\;\;\;\; g_{\phi\phi} = (\mathfrak{L}^{\prime})^{2-z} r^z + \beta  \mathfrak{L}' r+\alpha^2 (\mathfrak{L}')^2 + \frac{(\mathfrak{L}'')^2}{4(\mathfrak{L}^\prime)^2}\ee
where $\mathfrak{L} = \mathfrak{L}(\phi)$ and prime denotes a $\phi$ derivative. These black holes obey the tentative boundary conditions \eqref{bctent}. The Virasoro charges are all finite, integrable and conserved when $\xi$ has the form \eqref{chiass}. Explicitly, they are given by
\begin{equation}
\mathcal{Q}_L = \frac{z}{4\pi G  (2z-1)}  \int_0^{2 \pi} d\phi  \left[L(\phi) \left( \alpha^2 \mathfrak{L}'(\phi)^2 +{1\over 4}{\mathfrak{L}''(\phi){}^2 \over \mathfrak{L}'(\phi){}^2 }\right)
   + L'(\phi){\mathfrak{L}''(\phi) \over 2\, \mathfrak{L}'(\phi) }  \right]
\end{equation}
and $Q_{N} = 0$. The charge associated with $L_0$ is manifestly non-negative for all $z > 2$. The polynomial mode quadratic in time has nonzero charge $\mathcal{Q}_{N}$ but, as discussed for $z =2$, imposing regularity in the interior removes this mode. We therefore expect a chiral spectrum for all regular z-warped geometries as well.

When it is realized through the asymptotically conserved charges, the Virasoro algebra acquires a central extension given by:
\be\label{CentralCharges}
 c_z =  \frac{3  \ell  (1+ \mu \ell )}{2  G  \mu \ell}.
\ee
This value is the expected one for the Cardy formula to reproduce the entropy of z-warped black holes \re{zBHentropy}. The symplectic products between all modes obtained by linearizing \eqref{eq:cosmz} are exactly the one given in the previous section on null warped. The conclusions are therefore identical. We expect that an initial value problem will be well-posed with all unstable modes $\gamma$ set to zero.

\subsection*{Acknowledgements}
We thank Atish Dabholkar, Gaston Giribet, Niklas Johansson, Petr Horava, Gary Horowitz, Josh Lapan,  Charles Melby-Thompson, Sameer Murthy, Balt van Rees, Mukund Rangamani, Ashoke Sen, Andy Strominger and especially Don Marolf for many useful discussions. This research was  supported in part by the National Science Foundation under Grant No. PHY05-51164. D.A. was supported in part by DOE grant DE-FG02-91ER40654. The work of G.C. is supported in part by the US National Science Foundation under Grant No.~PHY05-55669, and by funds from the University of California. The works of S.dB. and S.D. are funded by the European Commission though the grants PIOF-GA-2008-220338 and PIOF-GA-2008-219950  (Home institution: Universit\'e Libre de Bruxelles, Service de Physique Th\'eorique et Math\'ematique, Campus de la Plaine, B-1050 Bruxelles, Belgium).

\appendix

\section{Geometry of null warped AdS$_3$}\label{nullgeom}

In this appendix, we briefly discuss some geometrical features of the null background geometry for both the Poincar\'e and the global patches.

\subsection{Geodesics}

We would like to study the behavior of geodesics in the null warped spacetime, with metric given by
\be
ds^2 = f(r) d \phi^2 + 2 r dt d\phi + \frac{dr^2}{4r^2}
\ee
where $\phi$ may or may not be identified. The case with
\be
f(r) = r^2 + \b r + \a^2 \;, \;\;\;\;\; \phi \sim \phi + 2 \pi
\ee
represents the black holes, whereas
\be
f(r) = r^2 -1\;, \;\;\;\;\; \phi \in \mathbb{R}
\ee
represents the global spacetime, as has been shown in \cite{Blau:2009gd}. The Poincar\'e patch is simply given by $f(r)=r^2$. We consider a geodesic with affine parameter $\l$
\be
r \frac{d\phi}{d\l} = - J \;, \;\;\;\;\; r \frac{dt}{d\l} + f(r) \frac{d\phi}{d\l} = -E
\ee
and $ds^2 = k d\l^2$
\be
k = \frac{1}{4r^2} \left(\frac{d r}{d\l}\right)^2 + \frac{2 J E}{r} - \frac{f(r) J^2}{r^2} \;, \;\;\;\;\; f(r) = r^2 +  \b r + a
\ee
We will mostly concentrate on the behavior of null geodesics, $k=0$. Letting $\tilde{\l} = 2 |J| \l$ and $\tilde{\b} = \half \b - \frac{E}{J}$, we find the solution\footnote{An additional solution exists for which $\l \r - \l$.}
\be
r(\l) = \half \, e^{\tilde{\l}} + \frac{\tilde{\b}^2-a}{2}  \, e^{-\tilde{\l}} - \tilde{\b} \;, \;\;\;\;\; \frac{dt}{d\l} = -\frac{E}{r} + \frac{J f(r)}{r^2}.
\ee
Several remarks are in place. First, unlike the case of Schr\"{o}dinger spacetimes studied in \cite{Blau:2009gd}, null geodesics can escape to infinity. The behavior as $r \r 0$ depends on the sign of $a$. If $a < 0$, then geodesics reach a minimum distance
\be
r_{min} = \sqrt{\tilde{\b}^2-a} -   \tilde{\b}  > 0. \label{minr}
\ee
Thus, the $r=0$ region is never reached because of the harmonic trap term $- d \phi^2$. We can therefore infer from the behaviour of geodesics that the boundary of the vacuum spacetime contains both $r=\infty$ and $r=0$. Note nevertheless that there exists a qualitative difference between the two, since the first can be reached by null geodesics in infinite proper time, whereas the second is never reached. There exists a minimum distance \eqref{minr} that null rays can attain in finite affine time.

If $a >0$, then null geodesics can reach $r<0$, and they do so in finite proper time. Nevertheless, in coordinate time
\be
\lim_{r\r \infty}t(r) \sim \pm J \ln r \;, \;\;\;\;\;\; \lim_{r\r 0} t(r) \sim  \mp \frac{J \sqrt{a}}{r}.
\ee
Thus, it takes infinite coordinate time to reach from $r$ finite to $r=0$, which indicates that $r=0$ is a horizon. Another way to see this is by noting that trajectories for which the future direction is given by decreasing $\tilde{\l}$ - namely, those geodesics for which $J <0$ - and that have $\tilde{\b}^2 < a$ always evolve towards more negative values of $r$.

\subsection{Null z-warped black holes and identifications}\label{proof}

In this section, we will show that there is no identification in the null z-warped background
\be
\frac{ds^2}{\ell^2} =  \rho^{z} d x_-^2 + 2 \rho \,dx_+dx_- + \frac{d\rho^2}{4\rho^2} \;, \;\;\;\;\; x^{\pm} \in (-\infty,\infty)   \label{Poincz}
\ee
for $z \neq 2$ that leads to the z-warped black holes
\be
\frac{ds^2}{\ell^2} = (  r^z+ \beta r + \alpha^2)\, d\phi^2 + 2 r \,dt d\phi + \frac{dr^2}{4r^2}  \label{BHcoordz}
\ee
where $\phi \sim \phi + 2\pi$ and $\alpha \neq 0$. In fact, we will show that, locally, one cannot write the background metric as \eqref{BHcoordz} when $z \neq 2$. Indeed, if this were the case, the two Killing vectors $\p_{\phi}$ and $\p_t$ of the metric \eqref{BHcoordz} should be expressible in terms of linear combinations of the three Killing vectors
\be
N = \partial_+ \, ,\quad N_1 = \partial_-\, , \quad N_0 = -\frac{2}{z}\rho\, \partial_\rho + x^- \partial_{-} + \frac{2-z}{z}x^+ \partial_+
\ee
of \eqref{Poincz} as
\be
\p_\phi =\alpha_0 N_0 + \alpha_{1}N_1+ \zeta N\;, \;\;\;\;\;
\p_t = \hat \alpha_0 N_0 + \hat \alpha_{1}N_1+\hat \zeta N \label{vecK}
\ee
for some \emph{constant} coefficients. Now, if both $\alpha_0$ and $\hat \alpha_0$ are zero, the change of coordinate is trivial and it does not map \eqref{Poincz} to \eqref{BHcoordz} when $\alpha \neq 0$. We will now assume $\alpha_0\neq 0$. The change of coordinates assuming $\hat \alpha_0\neq 0$ can be treated similarly with $\phi$ interchanged with $t$. The general solution of the first equation is given by
\be
\rho = F_1 (t,r)\, e^{-\frac{2\alpha_0}{z}\phi}\;, \;\;\;\;\;
x^-= F_2(t,r)\,e^{\alpha_0 \phi} - \frac{\alpha_{1}}{\alpha_0}\;, \;\;\;\;\;
x^+ = \frac{\zeta z}{\alpha_0(2-z)} \left( F_3(t,r)\, e^{\alpha_0 \frac{2-z}{z}\phi} - 1 \right).
\ee
Plugging these expressions into the second equation we find
\be
\p_t F_1 + \frac{2 \hat{\a}_0}{z} F_1 =0\;, \;\;\;\;\;\p_t F_2 - \hat{\a}_0 F_2 = \left(\hat{\a}_1-\frac{\hat{\a}_0 \,\a_1}{\a_0}\right) \,e^{-\a_0 \;\phi}
\ee
and
\be
\zeta \,\left(\p_t F_3 + \frac{z-2}{z} \hat{\a}_0 F_3\right) = \frac{2-z}{z} (\a_0 \, \hat{\zeta} - \hat{\a}_0 \,\zeta) \, e^{- \alpha_0 \frac{2-z}{z}\phi}.
\ee
Given that $\a_0 \neq 0$ the only way these equations can be satisfied is if
\be
\a_0 \, \hat{\a}_1 - \hat{\a}_0 \,\a_1 = (z-2)( \a_0 \, \hat{\zeta} - \hat{\a}_0 \,\zeta) = 0. \label{contradiction}
\ee
For $z = 2$, the second equation is trivially satisfied while for $z \neq 2$, it implies that $\hat \zeta = \frac{\hat \alpha_0 \zeta}{\alpha_0}$. As a consequence, the linear combinations appearing in \eqref{vecK} are equivalent and $\p_t$ is proportional to $\p_\phi$ (or is zero) which is absurd. For $z = 2$, we get rather that $\p_\phi = \alpha_0 N_0 + N $ while $\p_t = N $, in agreement with previous analyses.

\section{The linearized solution and its gauge-fixing}

\subsection{Details of the linearized solution}\label{solg12}

The equations of motion for $g_{1,2}^m$ that accompany \eqref{eqng3} are
\bea
\p_t g_2(r,t) & = & - 2 r^2 (4+3 \nu) g_3 - 4 r^3 \p_r g_3 - 3 t \nu C_1(r) - 3 \nu C_2(r) \non, \\
\p_t^2 g_1(r,t) &=& 72 \nu (\nu+1) r^4 g_3 + 48 r^5 \nu \p_r g_3 + 8 i m r^3 \p_t g_3 + 4 r^2 (r^z + \beta r + \a^2) \p_t^2 g_3 \non \\
&+& 36 r^2 \nu^2 [C_2(r) + C_1(r) t ] + 4 \nu C_3(r).
\eea

\subsection{Fixing the residual gauge}\label{gaugefixapp}

The temporal gauge condition $g_{\mu t}=0$ does not completely fix the gauge freedom. The residual gauge freedom is parameterized, for each fixed $m$, by three arbitrary functions of $r$, called $D_{1,2,3}^m(r)$:
\bea
\delta g_3^m (r,t) &=& \frac{r D_1'-D_1^m}{2 r^3} - 2 (r D_3'' + 2 D_3') \, t \non, \\
\delta g_2^m (r,t) &=& - \frac{i m D_1^m}{4 r^2} + r D_2' + (r^z+ \beta r + \a^2) \, D_3' + \left( \frac{D_1}{r} - D_1' + 2 i m r D_3'\right) \, t + 2r^2 (2 D_3' + r D_3'') \, t^2 \non, \\
\delta g_1^m (r,t) &=& (z r^{z-1} + \beta) D_1^m - 2 i m r D_2^m - 2 i m (r^z + \beta r + \a^2) D_3^m  + 2 im D_1^m t + 2 m^2 r D_3^m t \non, \\
&& \;\;\;\;\; \;\;\;\;\;\;\;\; - 4 r^3 (z r^{z-1} + \beta) D_3' t - 4 i m r^3 D_3' t^2
\eea
where for uncluttering we have dropped the superscript $m$ in presence of derivatives.

Due to the polynomial dependence on time of the above gauge transformations, we can consider that the gauge-fixing acts only on the particular (polynomial) solutions to the equations of motion. Below we will consider the $z=2$ and $z>2$ cases separately.

\subsection*{Gauge-fixing for $z=2$}

Let us first treat the case $z=2$, i.e. $\nu=1$. We are looking for a particular solution of the equation for $g_3^m$. Nevertheless, such a solution can always be gauged away by appropriately choosing $D_1^m$ and $D_3^m$. Thus, $g_3^m=0$ and this in turn implies that we can restrict the functions $C_i^m(r)$ to take the form
\be
C_1^m(r)=c_1^m \;, \;\;\;\;\; C_2^m(r) = c_2^m - \frac{c_3^m}{6r} - \frac{i m c_1^m }{4 r}\, (\ln r +1) \;, \;\;\;\;\; C_3^m(r) = c_3^m\, r + \frac{3 m c_1^m i}{2} \, r \ln r.
\ee
Next, we find the particular solutions for $g_{1,2}^m$. These depend on $c_{1,2,3}^m$ as well as other five arbitrary constants of integration. The leading behavior of the solution takes the form
\be
g_1^m(t,r) = 6 \,c_1^m r^2 t^3 + 18\, c_2^m t^2 r^2 + \ldots
\ee
Since we are not to consider perturbations that blow up as fast as the background, we have to set $c_1^m=c_2^m=0$, as well as gauge away two of the integration constants. In this case, the particular solution takes the form
\be
g_1^m(r,t) = \g^m \, r\, t + a^m + r \beta^m - c_3^m r t^2 + \frac{c_3^m \a^2}{4 r} + \frac{i c_3^m m}{2} \, t,
\ee
\be
g_2^m(r,t)= \frac{c_3^m t}{2 r}\;, \;\;\;\;\; g_3^m(r,t) =0.
\ee
The remaining gauge freedom is parameterized by the restricted functions
\be
D_1^m = d_1^m r\;, \;\;\;\;\; D_2^m = d_2^m - \frac{i m d_1^m}{4r} \;, \;\;\;\;\; D_3^m=d_4^m- \frac{d_3^m}{r}.
\ee
Requiring that the periodicity of the asymptotic metric remain unchanged sets
\be
d_1^m = i m d_4^m \;, \;\;\;\;\; d_3^m=0
\ee
and thus the remaining gauge freedom is
\be
\d g_1^m = - i m r (2 d_2^m + \beta d_4^m) - \frac{i m^3 d_4^m}{2} - 2 i m \a^2 d_4^m.
\ee
Using this remaining gauge freedom, we conclude that if $m \neq 0$ we can set $\beta^m$ and $a^m$ to zero by a coordinate transformation. In the case $m=0$, the solution is the same except that $\b^0, a^0$ are no longer removable by a gauge transformation.

\subsection*{Gauge-fixing for $z \neq 2$}

We follow the same procedure as in the previous subsection. We thus can gauge-fix $g_3^m=0$ and
\be
C_1^m(r)=  c_1^m r^{\frac{3(\nu-1)}{2}}  \;,\;\;\;\;\;C_2^m(r) =  c_2^m r^{\frac{3(\nu-1)}{2}} -\frac{c_3^m}{3 r (3 \nu-1)} - \frac{i m c_1^m}{6(\nu-1)} r^{\frac{3\nu-5}{2}}, \ee
\be
C_3^m(r) = r c_3^m + \frac{i m c_1^m}{\nu-1} r^{\frac{3\nu-1}{2}}.
\ee
The solution is the straightforward generalization of the one for $z=2$
\be
g_1^m=- \frac{2 c_3^m \nu r t^2}{3 \nu-1} + r \g^m t + \frac{i m \nu c_3^m}{3 \nu -1} t + \frac{c_3^m \nu }{2 (3 \nu -1)} r^{\frac{3 \nu-1}{2}} + r \beta^m + a^m + \frac{c_3^m \a^2 \nu}{2 r (3 \nu -1)},
\ee
\be
g_2^m = \frac{\nu c_3^m t}{(3\nu-1) r}, \quad g_3^m=0
\ee
where we have had again to set to zero $c_1^m, c_2^m$ etc. The diffeomorphisms that are still allowed must satisfy
\be
d_1^m = \frac{4 i m}{3\nu+1} d_4^m, \quad d_3^m=0,
\ee
with $d_2^m$ and $d_4^m$ arbitrary. The remaining diffeomorphisms are now:
\be
\d g_1^m = 2 m^2 r t d_4^m \left(1- \frac{2}{z} \right) + 2 i m r \left( d_4^m \b (\frac{1}{z}-1) -d_2^m\right) - \frac{i m^3}{z} d_4^m - 2 i m \a^2 d_4^m.
\ee
As before, the $\phi$-independent modes cannot be gauged away. An interesting observation is that in this case we can use the gauge freedom parameterized by $d_4^m$ to either gauge away the polynomial mode linear in time $\g^m$ or $a^m$ as before. The coefficient $\b^m$ can be gauged away for $m \neq 0$.

\section{Symplectic product}
\label{symplApp}

The symplectic product between two linearized graviton perturbations $h^{(1)}_{\mu\nu}$ and $h^{(2)}_{\mu\nu}$ around the background $\bar g_{\mu\nu}$ is defined as
\be
\langle h^{(1)} \arrowvert   h^{(2)}  \rangle  \equiv \Omega(h^{(1)} ,h^{(2)}) = -i \int_\Sigma \Omega[ h^{(1)}, h^{(2)*} ; \bar g ], \label{symplproduct}
\ee
while the symplectic norm of a linearized graviton perturbation $h^{(1)}_{\mu\nu}$  is
\be
\mathcal N \equiv \Omega(h^{(1)} ,h^{(1)}) = -i \int_\Sigma \Omega[ h^{(1)}, h^{(1)*} ; \bar g ],
\ee
where $\Omega$ is the symplectic structure and $\Sigma$ a hypersurface of constant $t$.  The symplectic structure is the sum of the symplectic structure for the Einstein-Hilbert action and the one of the gravitational Chern-Simons term. The standard covariant phase space symplectic structure for Einstein gravity can be found e.g. in \cite{Wald:1999wa}. Using covariant phase space methods, one can derive the Lee-Wald symplectic structure for the Chern-Simons term as
\be
\Omega[ \delta^{(1)}g , \delta^{(2)} g ; g ] = \delta^{(2)} \Theta (\delta^{(1)}g ;  g) - \delta^{(1)} \Theta (\delta^{(2)}g ;  g)
\ee
where the $2$ form $\Theta (\delta g ;  g)$ is given by (see also \cite{Andrade:2009ae} for a recent derivation)
\be
\Theta (\delta g ;  g) = \frac{1}{32\pi G \mu} (\Gamma_{\lambda\sigma}^\rho \delta \Gamma_{\rho \nu}^\sigma +2 R^\beta_{\;\; \nu} \delta g_{\lambda \beta} )\, dx^\lambda \wedge dx^\nu\, .
\ee

%\section{Asymptotic analysis}
\section{Asymptotic equations of motion}\label{asyexp}

For reasons explained in section \ref{ASG}, we have restricted our asymptotic analysis to metrics that only involve interger powers of $r$ in their asymptotic expansions. This is consistent with the equations of motion as explained below. We named the first coefficients in the asymptotic expansion with upper indices, which indicate at what level in the asymptotic expansion they appear in the equations of motion
\bea
g_{\phi\phi}(r,\phi,t) & = & r^2 + g_{\phi\phi}^{\sst{(1)}}(\phi,t) \, r +  g_{\phi\phi}^{\sst{(2)}}(\phi,t) + \O(r^{-1}) \non, \\
g_{\phi t}(r,\phi,t) & = & r + g_{\phi t}^{\sst{(1)}}(\phi,t) + \frac{1}{r}   g_{\phi t}^{\sst{(2)}}(\phi,t) + \O(r^{-2}) \non, \\
g_{t t}(r,\phi,t) & = &  \frac{1}{r} g_{\phi t}^{\sst{(1)}}(\phi,t) + \frac{1}{r^2}   g_{\phi t}^{\sst{(2)}}(\phi,t)+ \O(r^{-3}) \label{pertasym}
\eea
and similarily for the remaining components. It has been noticed that the null warped background ($\a=\b=0$) admits an anisotropic scaling symmetry of the coordinates \cite{Horava:2009vy}
\be
r \r \l r \;, \;\;\;\;\; t \r t \;, \;\;\;\;\; \phi \r \l^{-1} \phi. \label{anisc}
\ee
If we require that the perturbed metric \eqref{pertasym} be invariant under this symmetry, then the  ``metric coefficients'' $g_{\mu\nu}^{(i)}(\phi, t) $ must scale as
\be
g_{\mu\nu}^{(i)} (\phi, t) \r \l^i g_{\mu\nu}^{(i)} (\phi, t)
\label{metricconf}\ee
under \eqref{anisc} (see related considerations in \cite{Horava:2009vy}). This is an alternative definition of the index ``$(i)$''. The  $g^{(i)}_{\mu\nu}$ asymptotically determine the phase space \eqref{bndnaive}.

The usefulness of this definition stems from  the fact that if one  also expands the equations of motion \eqref{eomTMG} as a homogenous series in $r$, then each coefficient $E_{\mu\nu}^{(i)}$ depends only linearly on the metric $g^{(i)}_{\mu\nu}$ and its derivatives and non-linearly on $g^{(j)}_{\mu\nu}$, $j <i$ and its derivatives. This allows one to solve the equations of motion order by order.\footnote{If one decides, for example, to use the usual radial expansion that assigns the same order $i$ to all metric components that multiply a given power of $r$, she would find that she cannot solve for all functions at a given order, but would  have to wait for later orders in order to solve them.} We checked that the resulting equations admit consistent solutions up to third order in the expansion.

A considerable simplification occurs when one completely fixes the coordinates in radial gauge
\begin{eqnarray}
g_{rr} = \frac{1}{4r^2}, \qquad g_{r a} = 0,\quad a = \phi,\, t.
\end{eqnarray}
which can be achieved using an appropriate trivial diffeomorphism of the form \eqref{Triv}. We find that the most general asymptotic solution in our phase space up to second order has the form
\be
g^{(1)}_{\phi\phi} = -t^2 c(\phi)+  t \g(\phi)+\beta(\phi) \;, \;\;\;\;\; g_{t\phi}^{(1)}=\frac{1}{2} c(\phi)\;, \;\;\;\;\;\;g_{tt}^{(1)}=0 \label{asyg1}
\ee
while
\be
g^{(2)}_{\phi\phi} = -\frac{t^2}{2} \, c(\phi)^2 + \frac{t}{2}  \,  c(\phi) \g(\phi)+ \frac{t}{2}  \,  \partial_\phi \, c(\phi)+ a(\phi)
\ee
\be
g^{(2)}_{t\phi} = \frac{1}{16}\,c(\phi)^2 + \frac{5}{4} \,e(t,\phi)\;, \;\;\;\;\; g^{(2)}_{t\phi}=e(t,\phi)\label{asyg2}
\ee
where
\be
e(t,\phi)=e^{-2\sqrt{2}t} f_-^{(2%4
)}(\phi) + e^{2\sqrt{2}t} f_+^{(2%4
)}(\phi).
\ee
All functions of $\phi$ that appear are arbitrary. We recognise in the last line the exponential mode with $\om = 2 i \sqrt{2}$, which happens to be multiplied by an integer power of $r$. This time we found it from a (truncated) nonlinear asymptotic analysis.

\section{Birkoff-like theorem for null z-warped black holes}\label{birkhoff}

The statement we want to prove is that all stationary, axisymmetric solutions of TMG with two commuting Killing vectors one of which becomes null asymptotically are the null z-warped black holes. We will use the method of \cite{Maloney:2009ck} which is based on a rewriting of the TMG equations of motion presented in \cite{Clement:1994sb,Bouchareb:2007yx} and we will impose the null z-warped boundary conditions \eqref{bndnaive} or \eqref{bctent}.

\subsection{The equations of motion}

A general stationary axisymmetric solution of three-dimensional gravity reads
\be
\frac{ds^2}{\ell^2} = - (X^0+ X^1)\, d \phi^2 + 2 X^2 dt d\phi + (X^1-X^0)\, dt^2 + \frac{dr^2}{(X^1)^2 + (X^2)^2 -(X^0)^2}\label{Clform}
\ee
where $X^i=X^i(r)$ are only function of the radial coordinate and $\ell$ is a length scale which is related to the cosmological constant as $\Lambda = - \ell^{-2}$. Here $X^i$ is a vector in $\mathbb{R}^{1,2}$, with metric $\eta =$diag(${-1,1,1}$). We denote the norm of vectors $V^i$ as $\| V\|^2 = \eta_{ij} V^i V^j$. We also fix $\e_{012}=1$. The TMG equations of motion reduce to
\be
\ddot{X}_{\a}+ \frac{3}{2\mu}\, \e_{\a\b\gamma} \dot{X}^{\b} \ddot{X}^{\gamma} + \frac{1}{\mu} \, \e_{\a\b\gamma} X^{\b} \dddot{X}^{\gamma}  =0
\label{eqofm}
\ee
and the constraint equation
\be
\dot{X}^{\a} \dot{X}_{\a} - \frac{2}{\mu} \,\e_{\a\b\gamma} X^{\a} \dot{X}^{\b} \ddot{X}^{\gamma}+4\L=0. \label{constr}
\ee
We start with the once-integrated equations of motion of TMG, which read
\be
J^i =  \e^{i}{}_{j k} X^j \dot{X}^k -\frac{1}{2\mu} \left( 2 X^i X^m \ddot{X}_m - 2\ddot{X}^i X^m X_m -  X^i \dot{X}^m \dot{X}_m +  \dot{X}^i X^m \dot{X}_m \right) \label{eqmo}
\ee
where $J^i$ has been rescaled by a factor of $16 \pi G$ with respect to \cite{Bouchareb:2007yx}. Let us introduce the following scalar variables:
\be
\xi(r) = J_i X^i \;, \;\;\;\;\; \eta(r) = \e_{ijk} J^i X^j \dot{X}^k \;, \;\;\;\;\; z(r) = X^m X_m \equiv \|X\|^2.
\ee
Contracting the equations of motion \eqref{eqofm} in various ways we obtain
\be
2 \mu \xi = z \|\dot{X}\|^2 - \frac{\dot{z}^2}{4},
\ee
\be
\mu \eta = \mu (- z \|\dot{X}\|^2 + \frac{\dot{z}^2}{4}) + z \e_{ijk} X^i \dot{X}^j \ddot{X}^k,
\ee
\be
\dot{\eta} + \mu \dot{\xi} = \frac{\dot{z}}{4 z} (\eta + 2 \mu \xi).
\ee
Making use of the constraint equation \eqref{constr} it is not hard to show that
\be
\mu \xi + \eta = \frac{\dot{z}^2}{8} + 2 \L z.
\ee
Combining this with the previous equation we obtain\footnote{Provided that $\dot{z} \neq 0$.}
\be
\mu \xi = \ddot{z} z - \frac{\dot{z}^2}{8} + 6 \L z \;, \;\;\;\;\; \eta = - \ddot{z} z + \frac{\dot{z}^2}{4} - 4 \L z \;, \;\;\;\;\; \|\dot{X}\|^2 = 2 \ddot{z} + 12 \L. \label{xieta}
\ee
Finally, we obtain an extra equation, $F=0$, by contracting \eqref{eqmo} with $J^i$
\be
F = \mu \eta + \ddot{\xi} z - \frac{1}{4} \dot{\xi} \dot{z} - \half \xi \ddot{z} + \frac{3}{2} \xi \|\dot{X}\|^2 - \mu \|J\|^2 =\mu \eta + \ddot{\xi} z - \frac{1}{4} \dot{\xi} \dot{z} + \frac{5}{2} \xi \ddot{z} + 18 \L \xi - \mu \|J\|^2.  \label{jsqeom}
\ee
Substituting $\xi$, $\eta$ from  \eqref{xieta} into the above, we obtain one nonlinear differential equation for $z(r)$. Thus, if we could solve this equation, by using the definitions of $X^i$ in terms of $\eta, \xi, z $ we would obtain the most general solution  of  the TMG equations of motion which admits two commuting Killing vectors. In practice, the equation cannot be solved analytically, but we can use it to show that a given solution to the equations of motion is unique.

Before we proceed note that the following quantity
\be
E = (z \dot{\xi} - \half \dot{z} \xi)^2 - z \eta^2 +2 \mu \xi (\xi^2-\|J\|^2 z) \label{inteq}
\ee
satisfies
\be
2 z \dot{E} -3 \dot{z} E = 2 z (2 z \dot{\xi} - \xi \dot{z}) F = 0.
\ee
The solution to this equation is $E = \gamma z^{\frac{3}{2}}$, but it is not hard to show that $\g=0$ by re-expressing $\|\dot{X}\|^2$ in terms of $\xi,\eta,z$ and using various geometric relations. Since $E$ is more restrictive than \eqref{jsqeom}, we will be using it to prove the Birkhoff-like theorem.

\subsection{Asymptotically null z-warped black holes}
%{\bf several equations and comments changed}

The null warped black holes have
\be
X^0 = X^1 = -\half \,(r^z + \b \,r  +\a^2 )\;, \;\;\;\;\; X^2 = 2 r\;,\;\;\;\; \mu = 1- 2 z
\ee
where $z \geq 2$ should not be confused with the function $z(r)$ defined in the previous section. Compared to the main text, note also that the time $t$ differs by a factor 2 so that the form \eqref{Clform} holds and the epsilon tensor is defined with an opposite convention. The conserved charges of these spacetimes are
\be
J^i =\left(1-\frac{1}{\mu}\right) \a^2 (1,1,0)=\frac{2 z \a^2}{2z-1}(1,  1, 0) \;, \;\;\;\;\; \|J\|^2=0
\ee
which implies that the variables we are interested in are
\be
\xi = \frac{2 z \a^2}{2z-1} (X^1-X^0) \;, \;\;\;\;\; \eta = \xi \dot{X}^2 - \dot{\xi} X^2.\label{E16}
\ee
Taking $\L=-\ell^{-2}=-1$, the background solution simply has $\xi=\eta=0$, $z(r)= 4 r^2$ and trivially satisfies all the equations of motion. If the Killing vector $\p_t$ stopped being null in the interior of the spacetime, then one should obtain solutions where $\xi \neq 0$. After a trivial shift in $r$ so that $g_{rr} = \frac{1}{4r^2}+O(r^{-4})$, the null warped boundary conditions \eqref{bndnaive} or the null z-warped boundary conditions \eqref{bctent} imply that
\be
\xi \sim \O\left(\frac{1}{r}\right) \;, \;\;\;\;\;\eta \sim \O\left(\frac{1}{r}\right)\;, \;\;\;\;\;  z(r) \sim 4 r^2 + \O(1).\label{as15}
\ee
We now have to solve the equation of motion for $z(r)$. We assume an analytic expansion\footnote{It is probably not very hard to lift the assumption of analyticy, by finding the most general  solution to $E=0$ in \eqref{inteq}, e.g. numerically, which leads to a well-behaved metric. } of the form
\be
z(r) = 4 r^2 + \sum_{n=0}^{\infty} \frac{a_n}{r^n} \;, \;\;\;\;\; \eta(r) = \sum_{n=1}^{\infty} \frac{b_n}{r^n} \;, \;\;\;\;\;\xi(r)=\sum_{n=1}^{\infty} \frac{c_n}{r^n}. \label{expzex}
\ee
Comparing the left and right-hand side of $\eta$ and $\xi$ in \eqref{xieta} we obtain
\be
a_0 = 0,\qquad b_1 = -16 \,a_1,\qquad c_1 = \frac{12}{\mu}\,a_1.
\ee
Plugging the expansion into \eqref{inteq} we also find
\be
(\mu-3)(\mu +3) a_1 = 0.
\ee
If $z\neq 2$, then it follows that $a_1=0$. To show that $a_1=0$ also in the null warped case, we need a small extra piece of information, which is obtained by expanding
the expression
\be
|| \dot{X}||^2 = \frac{\p}{\p r}(-X^0+X^1)\frac{\p}{\p r}(X^0+X^1)+ (\dot{X}^2)^2
\ee
using the third relation in \eqref{xieta} and trading $-X^0+X^1$ for $\xi$ using \eqref{E16}. Given  that $X^0+X^1$ is nothing but $-g_{\phi\phi}$ and $X^2$ is $g_{t\phi}$, one can use the null warped boundary conditions as well as \eqref{as15} to show that  $a_1 = 0$. Then, plugging  \eqref{expzex} into \eqref{inteq} it follows that also all other $a_i=0$. Thus,
\be
z(r) = 4 r^2 \; \; \Rightarrow \;\; \xi(r)=0\;\; \Rightarrow \;\; g_{tt}=0 \;, \;\; g_{rr} = (2r)^{-2}.
\ee
As far as $g_{\phi\phi} =-( X^0 + X^1)$ is concerned, we can simply plug the above metric in the TMG equations of motion to find that the only solution is $g_{\phi\phi} = r^z + \b r + \a^2$. We have therefore proven the Birkoff-like theorem.

%\bibliographystyle{JHEP-2}
%\bibliography{master2}

\providecommand{\href}[2]{#2}\begingroup\raggedright\endgroup

\end{document}